\documentclass[11pt]{article}
\usepackage[nosort]{cite}
\usepackage{rotating}
\usepackage{footmisc}

\usepackage{amsmath}
\usepackage{amssymb}
\usepackage{amscd}
\usepackage{setspace}
\usepackage{bbm}
\usepackage{dsfont}
\usepackage{graphicx}
\usepackage{cancel}
\usepackage{relsize}

\def\R{\mathbb{R}}
\def\P{\mathbb{P}}
\def\C{\mathbb{C}}
\def\Q{\mathbb{Q}}
\def\Z{\mathbb{Z}}

\def\tr{\operatorname{tr}}

\def\dag{\dagger}
\def\rank{\operatorname{rank}}

\def\cA{\mathcal{A}}

\def\cE{\mathcal{E}}
\def\cF{\mathcal{F}}

\def\cL{\mathcal{L}}
\def\cU{\mathcal{U}}

\def\cO{\mathcal{O}}

\def\cQ{\mathcal{Q}}
\def\cY{\mathcal{Y}}

\def\cM{\mathcal{M}}
\def\2pii{(2 \pi i)}

\def\sig{\sigma}

\def\Xb{\bar{X}}

\def\cC{\mathcal{C}}

\def\cV{\mathcal{V}}

\def\rodland{R\o{}dland{}}
\def\zb{\bar{z}}
\def\wb{\bar{w}}
\def\cN{\mathcal{N}}
\def\mm{\mathfrak{m}}
\def\tsig{\widetilde{\sigma}}
\def\tmm{\widetilde{\mathfrak{m}}}
\newcommand{\rr}{{\mathsf{q}}}
\def\zzz{\xi}
\newcommand{\sss}{s}
\newcommand{\divisorsigma}{L} 

\newcommand{\be}{\begin{equation}}
\newcommand{\ee}{\end{equation}}
\newcommand{\bes}{\begin{equation*}}
\newcommand{\ees}{\end{equation*}}
\newcommand{\bea}{\begin{eqnarray}}
\newcommand{\eea}{\end{eqnarray}}
\newcommand{\beas}{\begin{eqnarray*}}
\newcommand{\eeas}{\end{eqnarray*}}

\newcommand{\p}{\partial}

\def\qvol#1{{\rm vol}_{q}( #1 )}
\def\slfrac#1#2{\left.\raisebox{0.5ex}{#1}\!\middle/\!\raisebox{-0.5ex}{#2}\right.}

\begin{document}
\numberwithin{equation}{section}
{
\begin{titlepage}
\begin{center}
\hfill BONN-TH-2012-21\\
\hfill NSF-KITP-12-150\\
\hfill UCSB Math 2012-33\\
\hfill IPMU 12-0165
\vskip 0.75in

{\Large \bf Two-Sphere Partition Functions and Gromov--Witten Invariants}\\

\vskip 0.4in

{ Hans Jockers${}^{a}$, Vijay Kumar${}^{b}$, Joshua M.~Lapan${}^{c}$, David R.~Morrison${}^{d,e}$, Mauricio Romo${}^{f}$}\\

\vskip 0.3in
{\relsize{-1}
\begin{tabular}{lll}
${}^{\, a}${\em Bethe Center for Theoretical Physics,} & ${}^{\, b}${\em KITP,}	&	${}^{\, c}${\em Department of Physics,} \\
$\phantom{{}^{\, a}}${\em Physikalisches Institut, Universit\"at Bonn} & $\phantom{{}^{\, b}}${\em University of California}	&	$\phantom{{}^{\, c}}${\em McGill University} \\
$\phantom{{}^{\, a}}${\em 53115 Bonn, Germany} & $\phantom{{}^{\, b}}${\em Santa Barbara, CA 93106, USA}	&	$\phantom{{}^{\, c}}${\em Montr\'eal, QC, Canada}		 \\[2ex]
\end{tabular}

\begin{tabular}{lll}
${}^{\, d}${\em Department of Mathematics,}  	& ${}^{\, e}${\em Department of Physics,} &	${}^{\, f}${\em Kavli IPMU (WPI),}\\
$\phantom{{}^{\, d}}${\em University of California} 				& $\phantom{{}^{\, e}}${\em University of California} & $\phantom{{}^{\, f}}${\em The University of Tokyo}\\
$\phantom{{}^{\, d}}${\em Santa Barbara, CA 93106, USA} 		&  $\phantom{{}^{\, e}}${\em  Santa Barbara, CA 93106, USA}	&	$\phantom{{}^{\, f}}${\em  Kashiwa, Chiba 277-8583, Japan}\\[2ex]
\end{tabular}
}

\end{center}

\vskip 0.35in

\begin{center} {\bf Abstract} \end{center}

Many $\cN=(2,2)$ two-dimensional nonlinear sigma models with Calabi--Yau target spaces admit ultraviolet descriptions as $\cN=(2,2)$ gauge theories (gauged linear sigma models).  We conjecture that the two-sphere partition function of such ultraviolet gauge theories --- recently computed via localization by Benini et al. and Doroud et al. --- yields the exact K\"ahler potential on the quantum K\"ahler moduli space for Calabi--Yau threefold target spaces. In particular, this allows one to compute the genus zero Gromov--Witten invariants for any such Calabi--Yau threefold without the use of mirror symmetry.  More generally, when the infrared superconformal fixed point is used to compactify string theory, this provides a direct method to compute the spacetime K\"ahler potential of certain moduli (e.g., vector multiplet moduli in type IIA), exactly in $\alpha'$.  We compute these quantities for the quintic and for R\o{}dland's Pfaffian Calabi--Yau threefold and find agreement with existing results in the literature.  We then apply our methods to a codimension four determinantal Calabi--Yau threefold in $\P^7$, recently given a nonabelian gauge theory description by the present authors, for which no mirror Calabi--Yau is currently known.  We derive predictions for its Gromov--Witten invariants and verify that our predictions satisfy nontrivial geometric checks.

\vfill

\noindent December 30, 2013

{\setlength\footnotemargin{0pt}
\let\thefootnote\relax\footnotetext{jockers@uni-bonn.de, vijayk@kitp.ucsb.edu, jlapan@physics.mcgill.ca, drm@math.ucsb.edu, \\mauricio.romo@ipmu.jp}
}

\end{titlepage}
}

\newpage

\section{Introduction}
\label{sec:overview}

Mirror symmetry
\cite{Dixon,LVW,CLS,Aspinwall:1990xe,Greene:1990ud}
is a valuable tool in understanding worldsheet quantum corrections to the moduli space of Calabi--Yau threefolds \cite{Candelas:1990rm}. This is because mirror symmetry maps the problem of computation of quantum corrections to a classical calculation in algebraic geometry. However, the technique is only applicable when the Calabi--Yau threefold under study has a known mirror construction. Barring a few exceptions, this is the case only for Calabi--Yau threefolds that have a realization as a complete intersection in a toric variety.

The  non-perturbative (in $\alpha'$) quantum corrections that one is interested in form a power series whose coefficients have come to be known as Gromov--Witten invariants \cite{MR809718, Dine:1987bq,Witten:1988xj}.  Roughly, these coefficients are related to the number of rational curves of fixed degree in the Calabi--Yau threefold. The method for evaluating Gromov--Witten invariants by using the mirror manifold,
pioneered in \cite{Candelas:1990rm,2param1,2param2,Hosono:1993qy,Hosono:1994ax},
has been shown to give accurate answers via an important ``Mirror Theorem'' in mathematics \cite{complete1,complete2} (again, this specifically applies to complete intersections in toric varieties).  The proofs of this mirror theorem have the interesting feature that they deal entirely with the curve-counting problem on the ``original'' Calabi--Yau manifold, using the mirror construction only for motivation.

In this paper, we conjecture an alternative way of computing Gromov--Witten invariants in physics which
 avoids the use of mirror symmetry. We consider the class of Calabi--Yau threefolds that can be realized via an ultraviolet gauge theory, constructed by means of the Gauged Linear Sigma Model (GLSM) \cite{Witten:1993yc}.  For abelian ultraviolet gauge groups, which generally lead to complete intersections in toric varieties, the fact that the ultraviolet theories contain enough information to evaluate Gromov--Witten invariants without using mirror symmetry was implicit in the early detailed studies of those theories \cite{Witten:1993yc,summing} that evaluated instanton expansions at high energy (as pointed out in the conclusions to \cite{summing}).  For nonabelian gauge groups, which generally lead to Calabi--Yau manifolds that are not complete intersections in toric varieties \cite{Witten:1993yc, Hori:2006dk, Donagi:2007hi, Hori:2011pd, Jockers:2012zr}, the situation has been a bit murkier.  Utilizing recent work in which the two-sphere partition function for such GLSMs was computed via localization \cite{Benini:2012, Doroud:2012}, we conjecture that this partition function computes the exact K\"ahler potential on the quantum K\"ahler moduli space of the Calabi--Yau threefold --- a quantity that can be used to extract the Gromov--Witten invariants.  We compute the partition function in two canonical examples (an abelian example, the quintic, and a nonabelian example, \rodland's~Pfaffian Calabi--Yau threefold in $\mathbb{P}^6$) and compare the Gromov--Witten invariants computed by our methods with those in the literature, finding perfect agreement in each case.  

Next, we use the conjecture to compute Gromov--Witten invariants for a Calabi--Yau threefold that can be realized as a codimension four determinantal variety in $\P^7$ --- a nonabelian GLSM for which was constructed by the present authors in \cite{Jockers:2012zr} --- for which a mirror has not yet been constructed.\footnote{The work of B\"ohm \cite{arXiv:0708.4402,arXiv:1103.2673} provides a promising proposal, but we have been unable to implement it well enough to produce a mirror for this example.}  We provide a table of these invariants and verify that they satisfy certain nontrivial geometric checks.
The result here is gratifying in that it involves a relatively simple computation in the high energy theory to produce the necessary ingredients for evaluating the Gromov--Witten invariants.

The outline of the paper follows: Section \ref{sec:review} provides an accessible account of special geometry and Gromov--Witten invariants; Section \ref{sec:conjecture} reviews the exact partition function on the two-sphere, states our main proposal, and explains how to extract Gromov--Witten invariants from the partition function; Section \ref{sec:examples} verifies our proposal in two examples with known mirrors; and Section \ref{sec:GN} contains our main new mathematical results, the Gromov--Witten invariants of a determinantal Calabi--Yau threefold in $\P^7$.  We end with a summary of our results and future directions.

\section{Review} \label{sec:review}
In this section, we review some aspects of $\cN=2$ special geometry, the moduli space of Calabi--Yau threefolds, and Gromov--Witten invariants.
The relevant object of study is the $\cN=(2,2)$ two-dimensional nonlinear sigma model (NLSM) with a Calabi--Yau threefold $Y$ as its target space.  The marginal deformations of the infrared superconformal theory can be identified with the complex structure moduli and complexified K\"ahler moduli of the Calabi--Yau threefold $Y$; in fact, the moduli space is locally a direct product of the complex structure moduli space $\cM_{CS}$ and the quantum-corrected K\"ahler moduli space $\cM_{\text{K\"ahler}}$, which are each local special K\"ahler manifolds governed by $\cN=2$ special geometry \cite{deWit:1984pk,MR717607,Strominger:1990pd,Candelas:1990pi,Bershadsky:1993cx,Craps:1997gp}. For readers already familiar with this topic, the relevant formula we will utilize is \eqref{eq:eK-flat}.

\subsection{Local special K\"ahler manifolds}

For a precise definition of local special K\"ahler manifolds, and for further details, we refer the reader to \cite{MR717607,Strominger:1990pd,Freed:1997dp}.\footnote{Local special K\"ahler manifolds are also often called projective special K\"ahler manifolds and are distinct from special K\"ahler manifolds --- see, e.g.,~\cite{Freed:1997dp}.} Part of the structure of a local special K\"ahler manifold $\cM$ of (complex) dimension $n$, with K\"ahler form $\omega$, includes a holomorphic line bundle $\cL$ as a subbundle of a holomorphic vector bundle $\cV$ of rank $n+1$ over $\cM$. The vector bundle $\cV$ comes with a flat connection $\nabla$ such that $\nabla\cL \subset \cV$, and the underlying real bundle $\cV_\R$ of real dimension $2n+2$ is equipped with a non-degenerate skew symmetric form $\langle \cdot , \cdot \rangle$ that extends to the complexification $\cV_\C$ of $\cV_\R$ of complex dimension $2n+2$.\footnote{This geometric structure gives rise to a Hodge filtration $\cF^3 \subset \cF^2 \subset \cF^1 \subset \cF^0$ of weight $3$, with $\cF^3\simeq \cL$, $\cF^2\simeq \cV$, $\cF^1\simeq \cL^\perp$, and $\cF^0\simeq \cV_{\C}$, where $\cL^\perp$ is the subspace of $\cV_\C$ perpendicular to $\cL$ with respect to the symplectic pairing $\langle \cdot , \cdot \rangle$ --- see, e.g.,~\cite{MR720288,MR1442525}.}

Locally, the K\"ahler potential $K$ of such a local special K\"ahler manifold $\cM$ --- with the K\"ahler form $\omega=\partial\bar\partial K$ --- is given in terms of a local holomorphic non-vanishing section $s$ of $\cL$ by
\begin{equation} \label{eq:Kgeneral}
  K \,=\,- \log \, i \langle s , \bar s \rangle \ .
\end{equation}
Note that the K\"ahler potential is only defined up to K\"ahler transformations $K \rightarrow K + f + \bar f$, where $f$ is a local holomorphic function. Such a K\"ahler transformation simply rescales the holomorphic section $s \rightarrow e^f s$.

The K\"ahler potential $K$ of the manifold $\cM$ can be conveniently expressed in terms of ``special projective coordinates'' $X^I$, $I=0,\ldots,n$, together with their conjugate special projective coordinates $\cF_I$, $I=0,\ldots,n$, as  \cite{deWit:1984pk,Strominger:1990pd,Freed:1997dp}
\begin{equation} \label{eq:Kspecial}
   K \,=\,  - \log \, i \left(  \Xb^I \cF_{I} -  X^I \bar\cF_I\right) \ .
\end{equation}
The coordinates $X^I$ and conjugate coordinates $\cF_I$ are also known as periods of the local special K\"ahler manifold $\cM$.

Furthermore, $\cN=2$ special geometry dictates that the conjugate periods $\cF_I$ --- viewed as functions of the periods $X^I$ --- are integrable to a holomorphic section $\cF(X^I)$ of the line bundle $\cL^{\otimes 2}$ \cite{deWit:1984pk,MR717607,Strominger:1990pd,Bershadsky:1993cx,Freed:1997dp}, which is homogeneous of degree two, i.e.,
\begin{equation}
   \cF_I \,=\, \frac{\partial\cF}{\partial X^I}  \ , \qquad \cF(\lambda X^I) = \lambda^2 \, \cF(X^I) \ .
\end{equation}
The section $\cF$ is called the $\cN=2$ prepotential of the local special K\"ahler manifold $\cM$.

\subsection{The complex structure moduli space $\cM_{CS}$}

The special K\"ahler manifold $\cM_{CS}$ describes a family of Calabi--Yau threefolds $\cY_\zzz$, where $\zzz$ is a local coordinate of some coordinate patch $U\subset \cM_{CS}$ parametrizing the $h^{2,1}(Y)$-dimensional complex structure moduli space of the Calabi--Yau threefold $Y$. Here the line bundle $\cL$ is identified with $H^{3,0}(\cY_\zzz)$ and the section $s$ with the holomorphic three-form $\Omega(\zzz)$.  Furthermore, we have the identifications $\cV_\R = H^3(\cY_\zzz,\R)$, $\cV_\C = H^3(\cY_\zzz,\C)$, and the symplectic pairing
\begin{equation}
    \langle \alpha , \beta \rangle \,=\, \int_Y \alpha \wedge\beta \ , \qquad \alpha,\beta \in H^3(\cY_\zzz,\C) \ .
\end{equation}
Then the K\"ahler potential~\eqref{eq:Kgeneral} of the complex structure moduli space $\cM_{CS}$ is locally given by
\begin{equation} \label{eq:Kperiods}
     K(\zzz,\bar\zzz) \,=\, -\log \, i \int_Y \Omega(\zzz) \wedge \overline{\Omega(\zzz)} \,=\,  - \log \, i \left(  \bar X^I \cF_I -  X^I \bar \cF_I\right) \ .
\end{equation}
The last expression involves the periods of $\Omega$,
\begin{equation} \label{eq:periods}
   \Pi(\zzz) \,=\, \left(X^I(\zzz),\cF_I(\zzz)\right)\,=\,\left(\int_{A^I} \Omega(\zzz), \int_{B_J} \Omega(\zzz) \right) \ , \qquad I,J=0,\ldots,h^{2,1} \ ,
\end{equation}
with respect to a canonical symplectic basis $(A^I,B_J)$ of $H_3(\cY_\zzz,\Z)$ satisfying
\begin{equation}
   \langle A^I , B_J \rangle = \delta^I_J \, , \qquad \langle A^I , A^J \rangle \,=\, \langle B_I , B_J \rangle = 0 \, .
\end{equation}

\subsection{The quantum K\"ahler moduli space $\cM_{\text{K\"ahler}}$}
The main player of this note is the quantum-corrected K\"ahler moduli space $\cM_{\text{K\"ahler}}$ of a Calabi--Yau threefold $Y$, which is defined as the corresponding space of chiral-antichiral and antichiral-chiral moduli of the underlying SCFT.  It is also a local special K\"ahler manifold, parametrizing the $h^{1,1}(Y)$-dimensional quantum K\"ahler moduli space of a family of Calabi--Yau threefolds $\cY_t$, where $t$ represent the complexified K\"ahler coordinates of $Y$ in some patch $U\subset \cM_{\text{K\"ahler}}$.

The vector bundle $\cV_\R$ corresponds to (the non-torsion part of) the K-theory group $K(Y)$. The K-theory group $K(Y)$ is generated by holomorphic vector bundles over $Y$ and the non-torsion part of $K(Y)$ is isomorphic to the non-torsion part of $H^{even}(Y,\Z)$ via the Chern character isomorphism
\begin{equation} \label{eq:chIso}
  {\rm ch}\! : ~ K(Y)\otimes_\Z \Q \overset\sim\longrightarrow H^{even}(Y,\Z) \otimes_\Z \Q \, ,~~ \cE \mapsto {\rm ch}(\cE)\ ,
\end{equation}
(see, for example, \cite{Moore:1999gb}).
For Calabi--Yau threefolds $Y$, there is a natural non-degenerate symplectic pairing on (the non-torsion part of) $K(Y)$ \cite{MR893604,Brunner:1999jq},
\begin{equation} \label{eq:Apair}
   \langle \, \cdot , \cdot \rangle \! : ~ K(Y) \times K(Y) \rightarrow \Z \, , ~~ (\cE,\cF) \mapsto \langle \, \cE , \cF \rangle :=  - \int_Y {\rm ch}(\cE \otimes \cF^*)\,{\rm Td}(Y) \ ,
\end{equation}
involving the Todd class ${\rm Td}(Y)$. This symplectic pairing naturally extends to the complexification $\cV_\C\simeq H^{even}(Y,\C)$, where we have identified $\cV_\C$ with $H^{even}(Y,\C)$ by the Chern character isomorphism \eqref{eq:chIso}.

In order to calculate the K\"ahler potential \eqref{eq:Kgeneral} using the symplectic pairing \eqref{eq:Apair}, we need to specify a section $s(t)$ of the bundle $\cL\subset \cV$,\footnote{The associated Hodge filtration and, hence, the definition of the bundle $\cL$ in the context of quantum K\"ahler moduli spaces is explained in~\cite{MR1442525}.} which is given by 
\cite{MR1442525,Mayr:2000as}
\begin{equation} \label{eq:sAmodel}
     s(t) \,=\,\frac{1}{\sqrt{{\rm Td}(Y)}} 
     \left({\bf 1} +  \sum_\ell \qvol{\cC^\ell}\,\omega_\ell +  \sum_\ell \qvol{\hat\cC_{\ell}}\, \hat\omega^\ell 
       + \qvol{Y}\, \omega^{(3)} \right) \ .
\end{equation}
Here the zero form ${\bf 1}$, the two forms $\omega_\ell$ and their Poincar\'e dual four forms $\hat\omega^\ell$, and the volume form $\omega^{(3)}$, generate (the non-torsion part of) $H^{even}(Y,\Z)$, while $\qvol{\, \cdot \,}$ calculates the complexified quantum volume of the algebraic cycles $(1,\cC^\ell,\hat\cC_{\ell},Y)$ of $H_{even}(Y,\Z)$ dual to the generators of $H^{even}(Y,\Z)$.

The quantum volumes in \eqref{eq:sAmodel} are expressed in terms of the complexified K\"ahler form $J(t)$ of the family of Calabi--Yau manifolds $\cY_t$ as
\begin{equation} \label{eq:qVolume}
   \qvol{\cC^\ell} \,=\,  \int_{\cC^\ell} J   \ , \qquad
   \qvol{\hat\cC_{\ell}} \,=\, \frac{1}{2!} \int_{\hat\cC_{\ell}} J\star J  \ , \qquad
   \qvol{Y} \,=\, \frac{1}{3!} \int_Y J\star J\star J  \ .
\end{equation}
Here `$\star$' indicates the product in the quantum cohomology ring $QH^{even}(\cY_t)$ \cite{Kontsevich:1994qz,MR1366548}, which reduces in the strict large volume limit to the wedge product `$\wedge$' in the topological ring $H^{even}(\cY_t)$. Then the section $s(t)$ is conveniently written as
\begin{equation}
     s(t) \,=\,\frac{1}{\sqrt{{\rm Td}(Y)}} \, \exp_\star J(t)  \ ,
\end{equation}
with the exponential function $\exp_\star$ taken in the quantum cohomology ring $QH^{even}(\cY_t)$.

Evaluating the K\"ahler potential \eqref{eq:Kgeneral} with the symplectic pairing \eqref{eq:Apair} and the section \eqref{eq:sAmodel}, we arrive at the expression
\bea \label{eq:AKone}
     K &=&  -\log\bigg[ - i\int_Y  \exp_\star\big( J(t) \big) \wedge \,\exp_\star\big(-\overline{J(t)}\big) \bigg]  \nonumber \\
&=& -\log\,  i \Big[ \overline{\qvol{Y}} - \qvol{Y} +  
     \sum_\ell \left(\qvol{\hat\cC_{\ell}}\,\overline{\qvol{\cC^\ell}} - \overline{\qvol{\hat\cC_{\ell}}}\, \qvol{\cC^\ell}  \right) \Big] \ ,
\eea
where $\overline{\qvol{\, \cdot \,}}$ is the complex-conjugate quantum volume. 

In general, it is difficult to explicitly determine \eqref{eq:qVolume} away from the large volume limit due to the relevance of quantum corrections in the product  structure of $QH^{even}(\cY_t)$.  In order to capture such quantum corrections, it is necessary to carry out a rather complicated localization computation in the virtual moduli space of stable maps from genus zero curves into $Y$ \cite{Kontsevich:1994na}.  However, when the family $\cY_t$ has a known mirror family $\check{\cY}_z$ with complex structure coordinates $z \in \cM_{CS}(\check Y)$, we can perform a classical computation in the topological $B$-model (complex structure moduli space of $\check Y$) and infer the quantum corrections in the $A$-model (quantum K\"ahler moduli space of $Y$). In this way, the exact quantum-corrected prepotential on $\cM_{\text{K\"ahler}}$ can indirectly be derived since mirror symmetry implies $\cM_{\text{K\"ahler}}(Y) = \cM_{CS}(\check{Y})$ \cite{Greene:1990ud,Candelas:1990rm}.   Our conjecture will provide an alternative method to determine (\ref{eq:qVolume}), valid even when no mirror is known.

Close to a large volume point, there is a distinguished choice of flat coordinates $t^\ell$, $\ell = 1, \ldots , h^{1,1}(Y)$, that provides an affine parameterization of the complexified K\"ahler form $J = \sum_\ell \omega_\ell t^\ell$ in terms of  an integral basis $\omega_\ell$ of $H^2(Y,\Z)$ whose generators lie within the closure of  the classical K\"ahler cone.\footnote{Such a large volume point is a singular point in the quantum K\"ahler moduli space: there, the flat connection $\nabla$ of $\cV$ has a regular singular point with maximally unipotent monodromy \cite{mirrorguide,delignelimit,compact}.}  In the vicinity of such a large volume point, the quantum volumes \eqref{eq:qVolume} take the form 
\begin{equation} \label{eq:qVolume2}
\begin{aligned}
     \qvol{\cC^\ell}\,&=\, t^\ell \ , \\
     \qvol{\hat\cC_\ell}\,&=\, \frac12  \sum_{m,n} \kappa_{\ell mn}t^mt^n + \sum_m a_{\ell m}t^m + b_\ell + \frac{\partial}{\partial t^\ell} F_{\textrm{Inst}}(t) \ , \\
     \qvol{Y}\,&=\, \frac{1}{3!} \sum_{\ell,m,n} \kappa_{\ell mn}t^\ell t^mt^n -b_\ell t^\ell+ \frac{i}{8\pi^3} c - 2 F_{\textrm{Inst}}(t) 
       +\sum_\ell t^\ell\frac{\partial}{\partial t^\ell}  F_{\textrm{Inst}}(t)  \ .
\end{aligned}
\end{equation}
Here, the coefficients $\kappa_{\ell mn}$ are the classical intersection numbers of the cycles $\hat\cC_\ell$, i.e.,
\begin{equation}
   \kappa_{\ell mn}\,=\, \#(\hat\cC_\ell \cap \hat\cC_m \cap \hat\cC_n) \,=\, \int_Y \omega_\ell\wedge\omega_m \wedge \omega_n \ .
\end{equation}
The coefficients $b_\ell$ and $a_{\ell m}$ are real and, up to monodromies, topological invariants of $Y$, as discussed in \cite{Candelas:1990rm,Brunner:1999jq,Mayr:2000as}. The coefficient $c$ is
\begin{equation} \label{eq:pert}
    c \,=\, \chi(Y) \,\zeta(3) \ ,
\end{equation}
where $\chi(Y)$ is the Euler characteristic of the Calabi--Yau threefold $Y$.  This correction can be traced to the only perturbative correction at the four-loop level of the underlying $\cN=(2,2)$ SCFT \cite{Grisaru:1986px,Candelas:1990rm}.  The remaining correction terms arise from worldsheet instanton corrections, which take the following form \cite{Aspinwall:1991ce}
\begin{equation} \label{eq:nonpert}
    F_{\textrm{Inst}}(t)\,=\, \frac{1}{(2\pi i)^3} \sum_{\substack{\eta\in H_2(Y,\Z) \\ \eta\ne 0}} N_\eta \, {\rm Li}_3 ( q^\eta) \ ,
\end{equation}
where
\begin{equation}
    {\rm Li}_k(q) \,=\, \sum_{n=1}^{+\infty} \frac{q^n}{n^k} \ , \qquad
    q^\eta = \exp \left( 2\pi i \int_\eta J \right)  = e^{2\pi i \sum_\ell \eta_\ell t^\ell} \ .
\end{equation}
The integers $N_\eta$ enumerate the genus zero worldsheet instanton numbers in the homology class $\eta$.  In this work, we refer to them as the integral genus zero Gromov--Witten invariants, in contrast to the rational Gromov--Witten invariants $n_\eta$ that are simply the coefficients of $q^\eta$ in the expansion of $(2\pi i)^3\,F_{\textrm{Inst}}(t)$.

Evaluating the K\"ahler potential \eqref{eq:AKone} around the large volume point using the explicit expressions for the quantum volumes \eqref{eq:qVolume2}, we obtain
\bea
\label{eq:eK-flat}
   e^{-K(t,\bar t)} &=&   -\frac{i}{6} \sum_{\ell,m,n}\kappa_{\ell mn} (t^\ell -\bar t^\ell)(t^m -\bar t^m)(t^n -\bar t^n) + \frac{\zeta(3)}{4\pi^3} \chi(Y) \\
    && \ + \frac{2i}{(2\pi i)^3} \sum_{\eta} N_\eta \Big( {\rm Li}_3(q^\eta) +{\rm Li}_3(\bar q^\eta) \Big)
       -\frac{i}{(2\pi i)^2}\sum_{\eta,\ell}N_\eta \Big( {\rm Li}_2(q^\eta) +{\rm Li}_2(\bar q^\eta) \Big)\eta_\ell  (t^\ell -\bar t^\ell) \ ,  \nonumber
\eea
where $\eta_\ell \,=\, \int_\eta \omega_\ell$.  Note that in the flat coordinates $t^\ell$, the K\"ahler potential~\eqref{eq:eK-flat} of the special K\"ahler manifold $\cM_{\text{K\"ahler}}$ is only determined up to K\"ahler transformations.

Let us remark on how we arrive at the form \eqref{eq:eK-flat} of the K\"ahler potential using mirror symmetry: in the vicinity of a large complex structure point, the $B$-model periods \eqref{eq:periods} of the mirror moduli space $\cM_{CS}(\check Y)$ take the characteristic form
\begin{equation} \label{eq:PB}
    \Pi \,=\, (X^0, X^\ell, \cF_\ell, \cF_0) \, = \, X^0\left( 1, \ t^\ell,  \ \tfrac{\partial}{\partial t^\ell}F, \ F_0 \right) \ , 
    \qquad \ell=1,\ldots,h^{1,1}(Y) \ .
\end{equation}
Here we have used the scaling properties of the prepotential to define $\cF(X^0, X^\ell) = (X^0)^2F(t)$ and $F_0= 2 F -\sum_\ell t^\ell \partial_\ell F$ in terms of the flat coordinates $t^\ell=\frac{X^\ell}{X^0}$. The holomorphic function $F(t)$, to which we also refer as the prepotential, now takes the form\footnote{Note that, in contrast to the K\"ahler potential, the prepotential \eqref{eq:Pre} is not a monodromy invariant quantity. As a consequence, the subleading classical terms are always only given up to monodromy transformations.}
\begin{equation} \label{eq:Pre}
  F(t) \,=\, \frac{1}{3!} \sum_{\ell,m,n} \kappa_{\ell mn} t^\ell t^m t^n + \frac12 \sum_{\ell,m} a_{\ell m} t^\ell t^m +
 \sum_\ell b_\ell t^\ell - \frac{i}{16\pi^3} c + F_{\textrm{Inst}}(t) \ .
\end{equation}
Finally, inserting these periods \eqref{eq:PB} into \eqref{eq:Kperiods} reproduces the K\"ahler potential~\eqref{eq:eK-flat}. This is also clear from recognizing that $\qvol{Y}=-F_0$ and $\qvol{\hat\cC_\ell} = \p_\ell F$.

\section{The partition function $Z_{S^2}$}
\label{sec:conjecture}
In this section, we will spell out the conjecture that the two-sphere partition function of \cite{Benini:2012, Doroud:2012} computes the quantum-corrected K\"ahler potential for the K\"ahler moduli space $\cM_{\text{K\"ahler}}$ of Calabi--Yau threefolds.  We then discuss the possible relationship of this conjecture to topological-anti-topological fusion \cite{Cecotti:1991me}.
Finally, we describe a procedure for how to use the conjecture to systematically extract the genus zero Gromov--Witten invariants from the two-sphere partition function. This sets the stage for Section \ref{sec:examples}, where we apply and check the presented approach for explicit Calabi--Yau threefold examples. 
The agreement of the Gromov--Witten invariants in these examples with existing results in the literature serves as strong evidence in favor of our conjecture.

\subsection{The conjecture $Z_{S^2} = e^{-K}$} \label{subsec:conjecture}
An $\cN=(2,2)$ globally supersymmetric field theory on flat euclidean space obeys the usual supersymmetry algebra with constant spinor parameters. If the theory possesses a vector R-symmetry, it was recently shown that one can place the theory on a two-sphere of radius $R$ while preserving both this R-symmetry and a global $\cN=(2,2)$ supersymmetry \cite{Benini:2012, Doroud:2012}. This requires a deformation of the Lagrangian of the theory by terms of order $\frac{1}{R}$ and $\frac{1}{R^2}$ (see \cite{Festuccia:2011ws} for a general discussion) and, correspondingly, the theory on $S^2$ obeys a deformed supersymmetry algebra with variational parameters given by conformal Killing spinors of $S^2$.  This deformation of the theory is distinct from a topological twist since the spinors remain sections of the spin bundle and, more to the point, the theory is not topological.
It is worth noting that, in general, there could be several deformations of a given supersymmetric theory on $\R^2$ that result in a supersymmetric theory on $S^2$.  The authors of  \cite{Benini:2012, Doroud:2012} made a particular choice that allowed them to exploit a fermionic symmetry to  localize the path integral (as in \cite{Witten:1988ze, Losev:1997tp, Lossev:1997bz, Nekrasov:2002qd, Pestun:2007rz}) and compute the exact partition function for $(2,2)$ gauge theories with charged matter on $S^2$.

Consider an $\cN=(2,2)$ GLSM with gauge group $U(1)^{\sss} \times G$, where $G$ is a product of simple Lie groups.  The matter content consists of chiral multiplets $\Phi_A$ in the irreducible representation $R_A$ of $G$, and charges $Q_A^\ell$ under the $\prod_{\ell=1}^\sss U(1)_\ell$ factor of the gauge group.  We include FI parameters $r_\ell$ and theta angles $\theta_\ell$ for each $U(1)$ factor in the gauge group and an appropriate gauge-invariant superpotential that describes a compact Calabi--Yau threefold $Y$; without loss of generality, we take the large volume regime of $Y$ to be $r_\ell \rightarrow +\infty$ for all $\ell$, and we recall that the Calabi--Yau condition is equivalent to the existence of a non-anomalous $U(1)_{\textsc{a}}$ axial R-symmetry.
For simplicity, we will assume that the FI parameters provide a rational basis for the second cohomology of $Y$ and, therefore, that $h^{1,1}(Y)=\sss$.\footnote{In general, there may be fewer FI parameters in the GLSM than there are complexified K\"ahler moduli on the Calabi--Yau threefold $Y$.  For instance, in the context of abelian GLSMs this happens if the generators of the K\"ahler cone of the Calabi--Yau threefold involve non-toric divisors. However, even in these cases we expect that the outlined procedure still yields the quantum-corrected K\"ahler potential for a subspace of the entire K\"ahler moduli space $\cM_{\text{K\"ahler}}(Y)$.\label{foot:FI}}
Taking the minimal $U(1)$ charge excitations to have charges $\pm 1$, then the theta angles will have $2\pi$ periodicities and the coordinates $z_\ell := \exp(-2\pi  r_\ell+i\theta_\ell)$, which respect this periodicity, will furnish good coordinates around the large volume point of $Y$, $z_\ell=0$.

The partition function for such a GLSM on $S^2$, deformed appropriately by 
$\cO(\frac{1}{R})$ terms and localized onto the Coulomb branch, takes the form  \cite{Benini:2012, Doroud:2012}\footnote{Note that the framework of \cite{Benini:2012, Doroud:2012} is more general, allowing one to compute the partition function in the presence of twisted masses and background gauge fields for the non-R flavor symmetries.}
\begin{equation}
Z_{S^2} = \frac{1}{|\mathcal{W}|} \sum_{\mathfrak{m}, \tmm } \int \left( \prod_{\mu=1}^{\rank(G)} \frac{d\sigma_{\mu}}{2\pi} \prod_{\ell=1}^\sss \frac{d\tsig_\ell}{2\pi} \right) Z_{\textrm{class}} (\tsig, \tmm )\ Z_{\textrm{gauge}} (\sigma, \mm)\ \prod_A Z_{\Phi_A} (\sigma, \tsig, \mm, \tmm)\, , \label{eq:ZS2-formula}
\end{equation}
where 
\begin{align}
Z_{\textrm{gauge}} & = \prod_{\alpha > 0} \left( \frac{(\alpha \cdot \mm)^2}{4} + (\alpha\cdot \sigma)^2 \right) \ ,\\
Z_{\Phi_{\!A}} & = \prod_{\rho \in R_A} \frac{\Gamma\Big(\frac{\rr[A]}{2} - i \rho \cdot \sigma - i \sum_\ell Q_A^\ell\, \tsig_\ell  - \frac{1}{2}\rho \cdot \mm -  \frac{1}{2}\sum_\ell Q_A^\ell\, \tmm_\ell\Big)}{\Gamma\Big(1-\frac{\rr[A]}{2}+i\rho \cdot \sigma+i \sum_\ell Q_A^\ell \, \tsig_\ell-\frac{1}{2}\rho\cdot \mm -\frac{1}{2}\sum_\ell Q_A^\ell\,  \tmm_\ell\Big)}\ , \\
Z_{\textrm{class}} & = \prod_\ell \, \exp\!\big(\!-4\pi i r_\ell \, \tsig_\ell  - i \theta_\ell\,  \tmm_\ell \big) \ .
\end{align}
The variables $\sigma$ and $\mm$ are vectors in the Cartan subalgebra of $G$ with $\sigma \in \R^{{\rm{rank}}(G)}$ a real vector, $\mm \in \Z^{{\rm{rank}}(G)}$ integral, and $\tsig_\ell$ and $\tmm_\ell$ parameterizing $\R^{\sss}$ and $\Z^\sss$. $\mathcal{W}$ denotes the Weyl group of $G$, and $|\mathcal{W}|$ its cardinality.
The expression for $Z_{\textrm{gauge}}$ involves a product over all the positive roots $\alpha$ of the simple factors of $G$ --- in the absence of any simple factors, this term is simply equal to one.
$Z_{\Phi_{\! A}}$ involves a product over all the weights $\rho$ of the representation $R_A$ of $G$, while $\rr[A]$ is the vector R-charge of the chiral multiplet $\Phi_{\! A}$.  The inner product $\cdot$ is the standard inner product on $\R^{\rm{rank} (G)}$. 
We will assume that the superpotential is sufficiently general so as to fix the R-charges of the various chiral multiplets up to mixing with $U(1)$ factors in the gauge group.

\noindent
We can now state our main conjecture:

\begin{quotation}
The partition function of a two-dimensional $\cN=(2,2)$ GLSM, defined on $S^2$ as in \cite{Benini:2012, Doroud:2012}, which flows in the infrared to a $\cN=(2,2)$ NLSM with a Calabi--Yau threefold as its target space, computes the K\"ahler potential $K(z,\bar z)$ of the quantum-corrected K\"ahler moduli space $\cM_{\text{K\"ahler}}$ according to
\begin{equation} \label{eq:Conj}
     Z_{S^2}(z,\bar z) \,=\, e^{-K(z,\bar z)} \ ,
\end{equation}
in terms of the introduced GLSM coordinates $z_\ell$.
\end{quotation}
In the remainder of this paper, we provide arguments in favor of this conjecture.

Note that the partition function \eqref{eq:ZS2-formula} is a real function of the complexified K\"ahler moduli $z_\ell$ and is invariant under monodromies (around the large volume point, generated by $\theta \rightarrow \theta + 2\pi$). The deformed theory on $S^2$ is such that the superpotential terms of the GLSM Lagrangian are $Q$-exact, where $Q$ is the fermionic charge with respect to which the localization is performed \cite{Benini:2012}.  As a consequence, the partition function $Z_{S^2}$ is independent of parameters in the superpotential, which correspond to complex structure deformations of the Calabi--Yau manifold that is described by the IR theory. We stress that this independence is a consequence of the choice of deformation of the flat space theory performed by \cite{Benini:2012, Doroud:2012}. This provides a zeroth order check on the proposal.

\subsection{Interpretation via $tt^*$ equations}


The authors of \cite{Benini:2012, Doroud:2012} evaluated the partition function using a second localization scheme that resulted in a different ``Higgs branch'' representation. Indeed, we should expect both representations to agree --- and they do --- since the partition function cannot depend on the choice of localization scheme used to calculate it (though it certainly can depend on the way in which the flat space theory is placed onto $S^2$). At least superficially, this Higgs branch representation allows for an interpretation in terms of  topological-anti-topological fusion \cite{Cecotti:1991me}: the partition function $Z_{S^2}$ is equal to the overlap of ground states in the $A$-twisted GLSM on a hemisphere, and an $\bar{A}$-twisted GLSM on the other hemisphere. This relationship, which we will now briefly sketch, may ultimately provide a proof of our conjecture. 
In three-dimensional gauge theories with $\cN=2$ supersymmetry, the partition function computed in  \cite{Kapustin:2009kz, Hama:2011ea} displays a similar ``factorization''. This fact was noted in \cite{Pasquetti:2011fj}, and further explored in \cite{Beem:2012mb}.

Consider an $\cN=(2,2)$ SCFT with marginal chiral operators $\mathcal{O}_\ell$ and marginal anti-chiral operators $\bar{\mathcal{O}}_{\bar{m}}$. We can deform the SCFT Lagrangian using these operators
\begin{equation}
   \cL \rightarrow \cL + \sum_\ell \left(\tau^\ell\int \, d^2\theta \ \mathcal{O}_\ell \,+\,c.c.  \right) \ ,
\end{equation}
while preserving superconformal invariance.  The $\tau^\ell$ serve as local coordinates on the moduli space of marginal deformations.  The (normalized) two-point correlation functions of marginal operators can be used to define the Zamolodchikov metric on the moduli space of marginal deformations \cite{Zamolodchikov:1986gt, Periwal:1989mx}
\begin{equation} \label{eq:g}
G_{\ell\bar{m}}=(\langle \mathds{1} \rangle_{S^2})^{-1} \lim_{x\rightarrow \infty} x^4 \, \langle\bar{\mathfrak{o}}_{\bar{m}}(x) \ \mathfrak{o}_{\ell}(0)\rangle_{S^2} \, ,
\end{equation}
where $\mathfrak{o}_\ell = \mathcal{O}_\ell \vert_{\theta=\bar{\theta}=0}$.

In the $tt^*$ construction, described in \cite{Cecotti:1991me}, the authors studied the twisted SCFT ($A$ or $B$ twist) on a hemisphere with a chiral operator inserted at the pole. The hemisphere path integral of the twisted theory defines a state-operator correspondence between chiral operators inserted at the pole and supersymmetric ground states (up to the addition of $Q$-exact states) on the boundary $S^1$.  The topological twist allowed them to deform the hemisphere into an infinitely long cigar geometry, producing the unique ground state $\vert \ell\rangle$ corresponding to the operator $\mathcal{O}_\ell$ by projecting out any $Q$-exact pieces.  Performing the anti-topological twist on another hemisphere and constructing the anti-topological ground states $\vert \bar{m} \rangle$, corresponding to the anti-chiral operators $\bar{\mathcal{O}}_{\bar{m}}$, allowed the authors to give another useful interpretation of the Zamolodchikov metric in terms of the overlaps of these ground states:
\begin{equation}
G_{\ell\bar{m}} = \frac{\langle \ell\vert \bar{m} \rangle}{\langle 0 \vert \bar{0}\rangle}\, .
\end{equation}
Here the state $\vert 0 \rangle$ corresponds to the unit operator in the chiral ring. Moreover, the metric $G_{\ell\bar{m}}$ can be derived from a K\"ahler potential defined through
\begin{equation}
e^{-K} = \langle 0 \vert \bar{0} \rangle\, .
\end{equation}
Thus, in a $(2,2)$ SCFT corresponding to a Calabi--Yau threefold, we can obtain information about the K\"ahler potential of the quantum K\"ahler moduli space from the $A$-twist and that of the complex structure moduli space from the $B$-twist.  Since the $A$-twist ($B$-twist) can be carried out in any $(2,2)$ supersymmetric theory with non-anomalous vector (axial) R-symmetry, the $tt^*$ construction actually extends many concepts of special geometry away from the superconformal point.

Consider, now, the $A$-twist of the GLSM described in Section \ref{subsec:conjecture} on a hemisphere with the unit operator inserted at the pole. This corresponds to the ground state we labeled $\vert 0\rangle$. The path integral over the hemisphere, by the localization argument, receives contributions from vortex configurations that satisfy \cite{summing} 
\begin{equation}\label{vortexeqs}
 -r_\ell+\sum_A Q_A^\ell \tr(\phi_A\phi_A^{\dag}) =F^\ell_{12} \ , \qquad  D_{\bar{z}}\phi_A=0 \,  \ .
\end{equation}
Here we have assumed that the FI parameters $r_\ell$ have been tuned such that the gauge group is completely broken.\footnote{This may not be true in general. It could happen that a nonabelian group is unbroken for all values of the FI parameter.} Hence, the scalar fields in the vector multiplets are set to zero.  The equations \eqref{vortexeqs} are exactly the same as the equations satisfied by configurations that contribute to the partition function in the Higgs branch representation in \cite{Benini:2012, Doroud:2012} (this observation was also made in \cite{Doroud:2012}). Moreover, the localization equations on the other pole reduce to the anti-vortex equations, corresponding to the $\overline{A}$-twist. This suggests a connection with $tt^*$ fusion, although it is possibly only superficial (for example, the fermions here have antiperiodic boundary conditions around the equator, whereas we would expect periodic boundary conditions for $tt^*$ fusion).

A proof of our conjecture may be possible if we can show that the quantity $\langle 0 \vert \bar{0} \rangle$ is precisely computed by the partition function of the GLSM on $S^2$, as defined in \cite{Benini:2012, Doroud:2012}. This would require studying the vortex configurations of the $A$-twisted GLSM in the presence of boundaries with the appropriate boundary conditions. It would be interesting to investigate whether the conjecture that $Z_{S^2} = \langle 0 \vert \bar{0} \rangle$ holds away from conformality, i.e., when the axial R-symmetry is anomalous \cite{Cecotti:1991vb}.   We leave this to future work.



\subsection{Extracting Gromov--Witten invariants from the partition function}
\label{sec:procedure}
We now explain how, with the Euler characteristic $\chi(Y)$ as additional input, the conjecture \eqref{eq:Conj} can be used to extract the Gromov--Witten invariants from the partition function $Z_{S^2}$.  Alternatively, the agreement we find between computations in Section \ref{sec:examples} and known results for Gromov--Witten invariants can be thought of as strong evidence for our conjecture.
For ease of exposition, in this section we will assume that a large volume point is located at $z_\ell = 0$.  To bring the partition function $Z_{S^2}(z,\bar{z})$ into the normal form \eqref{eq:eK-flat} and to extract the Gromov--Witten invariants, we use the following algorithm:

\begin{enumerate}
\item Evaluate $Z_{S^2}(z,\bar{z}) = e^{-K}$ by contour integration as an expansion around large volume;

\item Isolate the perturbative $\zeta(3)$ term and perform a K\"ahler transformation $K = K' + X^0(z) + \overline{X^0(z)}$ in order to reproduce the constant term $\frac{\zeta(3)}{4\pi^3} \chi(Y)$ in \eqref{eq:eK-flat};\footnote{Alternatively, if the classical intersection numbers associated to the complexifed FI parameters $i r_\ell+\frac{\theta_\ell}{2\pi}$ are known, we can read off the same K\"ahler transformation $X^0(z)$ from the coefficients of the $\log^3\bar{z}$-terms, which may be computationally simpler to obtain.}

\item Read off the holomorphic part of the coefficient of $\log \zb_m \log \zb_n$, which should then be identified with
\begin{equation}
  -\frac{i}{2(2\pi i)^2} \kappa_{\ell mn} t^\ell \ .
\end{equation}
Use this to extract the flat coordinates $t^\ell$, which must have the form
\begin{equation} \label{eq:mmap}
  t^\ell = \frac{\log z_\ell}{2\pi i} + t_{(0)}^\ell + f^\ell(z)\ ,
\end{equation}
where $f^\ell(z)$ is a holomorphic function satisfying $f^\ell(0)\in i\mathbb{R}$ and $t_{(0)}^\ell \in [0,1)$ --- this determines the ``mirror map'' up to the undetermined constants $t_{(0)}^\ell$;

\item Invert the ``mirror map'' \eqref{eq:mmap} to obtain the $z_\ell$ as a function of $t^\ell$,
\begin{equation}
z_\ell = e^{-2\pi i t_{(0)}^\ell} \big( q_\ell + O(q^2)\big)\, ,
\end{equation}
where $q_\ell := e^{2\pi i t^\ell}$\,;

\item Fix the $t_{(0)}^\ell$ by demanding the lowest order terms in the instanton expansion be positive; and, finally,

\item  Read off the rational Gromov--Witten invariants from the coefficients in the $q$-expansion --- the integral genus zero Gromov--Witten numbers (roughly, the ``number of rational curves'') can then be obtained by the usual multi-covering formula (\ref{eq:nonpert}).

\end{enumerate}

\section{Examples}
\label{sec:examples}

In this section, we explicitly compute the Gromov--Witten invariants for certain Calabi--Yau threefolds using the partition function, as outlined above. We begin with the familiar example of the quintic hypersurface in $\P^4$, whose quantum K\"ahler moduli space was first analyzed using  mirror symmetry in \cite{Candelas:1990rm}. Next, we consider the Pfaffian Calabi--Yau threefold in $\P^6$ whose Gromov--Witten invariants were first computed by \rodland{} using mirror symmetry \cite{Rodland9801} and later studied via a GLSM constructed in \cite{Hori:2006dk, Hori:2011pd}.  We compute the partition function of this GLSM and find that the Gromov--Witten invariants determined by our procedure agree with those computed in \cite{Rodland9801, Tjotta1999}.

\subsection{The quintic threefold}

A degree $n$ hypersurface in $\P^{n-1}$ is described by a GLSM with gauge group and matter content summarized in Table \ref{tab:quintic}.  The $U(1)_{\textsc{v}}$ R-charge is kept partially arbitrary in terms of a parameter $\rr$.  The theory also has a superpotential
\begin{equation}
W = P G_n(\Phi)\, ,
\end{equation}
where $G_n(\Phi)$ refers to a generic homogeneous degree $n$ polynomial.
\begin{table}[t]
\begin{tabular}{|c||c|c|}
\hline
Field 	& $U(1)$   & $U(1)_{\textsc{v}}$  \\
\hline
\hline
$\Phi_a$ 	& $+1$	& $2\rr$	 \\
$P$		& $-n$	& $2-2n\rr$	 \\
\hline
\end{tabular}
\centering
\caption{Gauge group and matter content of a GLSM describing a Calabi--Yau hypersurface in $\P^{n-1}$. The index $a = 1,\ldots, n$.}
\label{tab:quintic}
\end{table}
The model has two phase descriptions depending on the value of the Fayet-Iliopoulos parameter $r$, with $r \gg 0$ describing a nonlinear sigma model phase and $r \ll 0$ a Landau-Ginzburg orbifold phase \cite{Witten:1993yc}. The exact partition function for this hypersurface can be written down for this model using the formulas in \cite{Benini:2012, Doroud:2012} as
\begin{equation}
Z_{\textrm{hyp}} = \sum_{m\in \Z} e^{-i \theta m} \int_{-\infty}^\infty \frac{d\sigma}{2\pi} \, e^{-4\pi i r \sigma} \left( Z_\Phi \right)^n Z_P \ ,
\end{equation}
where
\begin{equation}
Z_\Phi = \frac{\Gamma\big(\rr-i \sigma - \frac{m}{2}\big)}{\Gamma\big(1-\rr+i\sigma-\frac{m}{2}\big)}\ , \qquad Z_P = \frac{\Gamma\big(1-n\rr + n i\sigma + \frac{n m}{2}\big)}{\Gamma\big(n\rr - n i\sigma +  \frac{n m}{2}\big)}\ .
\end{equation}
A division between two sets of poles in the partition function suggests  we choose $0<\rr<\frac{1}{n}$ (see Appendix \ref{sec:quintic-calc}).  This  also happens to correspond to non-negative R-charges, simplifying the computations in \cite{Benini:2012, Doroud:2012}, and so we restrict ourselves to this range.  

It is convenient to change variables to $\tau = \rr - i\sigma$, in terms of which, the partition function is
\begin{equation}
Z_{\textrm{hyp}} = e^{-4\pi  r\rr}\sum_{m\in \Z} e^{-i \theta m} \int_{\rr-i\infty}^{\rr+i\infty} \frac{d\tau}{2\pi i} \, e^{4\pi  r \tau}  \frac{\Gamma\big(\tau -\frac{m}{2}\big)^n}{\Gamma\big(1-\tau-\frac{m}{2}\big)^n}\, \frac{\Gamma\big(1-n\tau + \frac{n m}{2}\big)}{\Gamma\big(n\tau +  \frac{n m}{2}\big)} \ . \label{eq:quintic-pf-formula}
\end{equation}
The integral in equation \eqref{eq:quintic-pf-formula} is easily evaluated by the method of residues, with the way in which we close the contour clearly dependent on the sign of $r$.  When $r \gg 0$, the contour can be closed in the left half-plane yielding the answer (see Appendix \ref{sec:quintic-calc} for the details and for the expansion around the Landau--Ginzburg orbifold point)
\begin{gather}
Z_{\textrm{hyp}} =(z\zb)^\rr   \oint \frac{d\epsilon}{2\pi i}  (z\zb)^{-\epsilon} \ \frac{\pi^{n-1} \sin(n\pi\epsilon) }{\sin^n (\pi \epsilon)} \left\vert \sum_{k=0}^\infty  (-1)^{nk} z^k
 \frac{\Gamma(1+nk-n\epsilon) }{\Gamma(1+k-\epsilon)^n}\right\vert^2 \, ,
\end{gather}
where the contour of integration only encircles the pole at the origin $\epsilon=0$, the complex conjugation does not act on $\epsilon$, and where $z := \exp(-2\pi r + i \theta )$.  For the quintic ($n=5$), we have
\begin{gather}
Z_{\textrm{quintic}} =(z\zb)^\rr \oint \frac{d\epsilon}{2\pi i}   (z\zb)^{-\epsilon} \ \frac{\pi^{4} \sin(5\pi\epsilon) }{\sin^5 (\pi \epsilon)} \left\vert \sum_{k=0}^\infty  \left(-z\right)^k
 \frac{\Gamma(1+5k-5\epsilon) }{\Gamma(1+k-\epsilon)^5}\right\vert^2 \, . \label{eq:quintic-LV}
\end{gather}
Having required $0<\rr<\frac{1}{5}$, notice that the remaining dependence on $\rr$ is only through an overall multiplicative factor that can be removed by a K\"ahler transformation.  In what follows, we will disregard the pre-factor $(z\zb)^\rr$ by taking $\rr\to \frac{1}{5}^-$ (this is the natural choice of R-charge here since this model has a Landau-Ginzburg phase where $P$ obtains a \textsc{vev}, but for models without Landau-Ginzburg phases it is less clear from the UV theory how one should choose $\rr$).

We now demonstrate that the Gromov--Witten invariants, as determined by the procedure of Section \ref{sec:procedure}, agree with those computed in \cite{Candelas:1990rm}. First we extract the coefficient of $\zeta(3)$ in \eqref{eq:quintic-LV}, which determines the K\"ahler transformation $X^0(z)$ to be performed.  After performing the K\"ahler transformation, the K\"ahler potential becomes
\begin{equation}
e^{-K'} = -\frac{1}{8\pi^3} \frac{Z_{\textrm{quintic}}}{X^0(z) \overline{X^0(z)}}\ ,
\end{equation}
where
\begin{equation}
X^0(z) = \sum_{k=0}^\infty \frac{\Gamma(1+5k)}{\Gamma(1+k)^5}\ (-z)^k\ .
\end{equation}
It is interesting to observe that $X^0(z)$ is precisely the ``fundamental period'' of the quintic as determined by mirror symmetry ($z$ is rescaled by a factor of $-5^5$ relative to the formulas in \cite{Candelas:1990rm}). This suggests that our methods are closely related to toric mirror symmetry
\cite{Batyrev1993,Batyrev-Borisov-1994}, in which the periods are known
\cite{BvS} to be generalized hypergeometric functions \cite{GKZ2}.

Next, we determine the mirror map through the coefficient of the $\log^2 \zb$ term, which yields
\begin{equation}
t = t_{(0)} + \frac{1}{2\pi i} \left( \log z -770\, z + 71 7\,825\, z^2 + \ldots \right)\, .
\end{equation}
Inverting the mirror map, we find that the leading instanton correction is exactly $-2875 e^{-2\pi i t_{(0)}}$, fixing the undetermined constant $t_{(0)}\in [0,1)$ to be $t_{(0)}= \frac{1}{2}$.  With this choice, the integral genus zero Gromov--Witten invariants are
\begin{equation}
2\,875\, , \ \ 609\,250\, , \ \ 317\, 206\, 375\, , \ \ 242\, 467\, 530\, 000\, , \ \ \ldots \ ,
\end{equation}
and agree with the numbers in the literature \cite{Candelas:1990rm}.

\subsection{\rodland's Pfaffian Calabi--Yau threefold}

In this section, we analyze the partition function of a Calabi--Yau subvariety of $\P^6$ defined by the rank 4 locus of a $7\times 7$ antisymmetric matrix whose entries are linear in the homogeneous coordinates of $\P^6$.  \rodland{} conjectured that this Calabi--Yau threefold is in the same K\"ahler moduli space as a complete intersection of seven hyperplanes in the Grassmannian $G(2,7)$ \cite{Rodland9801}, and Hori and Tong later gave a ``physics proof'' of this conjecture by constructing a GLSM that reduces to a nonlinear sigma model on the two Calabi--Yau threefolds for two different limits of the FI parameter \cite{Hori:2006dk}.  The corresponding GLSM has gauge group $U(2)$, matter content as in Table \ref{tab:rodland}, and superpotential
\be
W = \tr\big(A(\Phi) P^T \varepsilon P\big) = A^a_{ij} \Phi^a P_{\alpha j} \varepsilon_{\alpha\beta} P_{\beta i} \, .
\ee
Here $\alpha,\beta=1,2,$ are fundamental $U(2)$ indices and $\varepsilon$ is a $2\times 2$ antisymmetric matrix with $\varepsilon_{12}=1$. The seven antisymmetric $7\times 7$ matrices $A^a_{ij}$ serve as defining matrices for the two Calabi--Yau threefolds described by the GLSM, determining their complex structures.  A division between two sets of poles in the partition function, as in the quintic analysis, suggests we choose $0<\rr<\frac{1}{2}$, which again happens to correspond to non-negative R-charges.
\begin{table}[t]
\begin{tabular}{|c||c|c|}
\hline
Field 	& $U(2)$   & $U(1)_{\textsc{v}}$  \\
\hline
\hline
$\Phi_a$ 	& $1_{-2}$	& $2-4\rr$	 \\
$P_i$		& $\Box_{+1}$	& $2\rr$		 \\
\hline
\end{tabular}
\centering
\caption{Gauge group and matter content of the Hori-Tong GLSM. The subscript denotes the charge under $\det(U(2))$. Here $i,a = 1,\ldots, 7$.}
\label{tab:rodland}
\end{table}
Following the methods of \cite{Benini:2012, Doroud:2012}, the partition function is given by
\bea
\label{eqn:RodZ}
Z &=& \frac{1}{2} \sum_{m_0,m_1\in\mathbb{Z}} \int_{i\rr-\infty}^{i\rr+\infty}\frac{d\sigma_0}{2\pi}\int_{i\rr-\infty}^{i\rr+\infty} \frac{d\sigma_1}{2\pi} \Big[ \big(\tfrac{m_0-m_1}{2}\big)^2 + (\sigma_0 - \sigma_1)^2 \Big] \Bigg[ \frac{\Gamma(-i\sigma_0 - \tfrac{m_0}{2})}{\Gamma(1+i\sigma_0 - \tfrac{m_0}{2})}    \nonumber \\
&& \qquad\quad \times \frac{\Gamma(-i\sigma_1 - \tfrac{m_1}{2})}{\Gamma(1+i\sigma_1 - \tfrac{m_1}{2})} \frac{\Gamma(1+i\sigma_0 + i\sigma_1 + \tfrac{m_0+m_1}{2})}{\Gamma(-i\sigma_0-i\sigma_1+\tfrac{m_0+m_1}{2})} \Bigg]^7 \, e^{-4\pi i r(\sigma_0+\sigma_1)-i\theta(m_0+m_1)- 8\pi \rr r} \ .  \qquad \quad
\eea

\noindent \underline{\bf Grassmann Phase: $r \gg 0$}

In the $r \gg 0$ phase, the GLSM flows to a nonlinear sigma model whose target space is the complete intersection of seven hyperplanes in the Grassmannian $G(2,7)$. The partition function can be evaluated in a similar manner to the quintic example, closing the contour in the lower half plane for both $\sigma_0$ and $\sigma_1$, and is given by
\bea
\label{eqn:Rod-Grass}
Z &=& \frac{(z\zb)^{2\rr}}{2} \oint \frac{d^2\epsilon}{(2\pi i)^2} \frac{\pi^7\sin^7\pi(\epsilon_0+\epsilon_1)}{\sin^7\pi\epsilon_0 \, \sin^7\pi\epsilon_1}  (z\bar{z})^{\epsilon_0+\epsilon_1}
   \nonumber  \\
&& \qquad\qquad \times \left\vert \sum_{K= 0}^\infty (-z)^{K}  \sum_{k=0}^K  (2k-K + \epsilon_0-\epsilon_1) \frac{ \Gamma(1+K+\epsilon_0+\epsilon_1)^7 }{\Gamma(1+k+\epsilon_0)^7\Gamma(1+K-k+\epsilon_1)^7} \right\vert^2  \, ,  \qquad\quad
\eea
where the integrals over $\epsilon_{0,1}$ are performed along contours only enclosing the poles at $\epsilon_{0,1}=0$ and where complex conjugation does not act on $\epsilon_{0,1}$.  Again, we use a K\"ahler transformation to remove the dependence on the parameter $\rr$.

Determining the Gromov--Witten invariants in this phase proceeds exactly as in the quintic example: we first read off the K\"ahler transformation from the $\zeta(3)$ term, which takes the form
\bea
\label{eqn:OurRodPeriod}
X^0_{\textrm{Grass}}(z) &=& -\frac{1}{2} \sum_{K=0}^\infty \sum_{k=0}^K {K \choose k}^7 \Big[ -2 + 7(K-2k) \big( H_{K-k} - H_k \big) \Big] (-z)^K   \\
&=& 1 + 5\,z + 109\, z^2 + 3\, 317\, z^3 + 121\,501\, z^4 + 4\,954\,505\, z^5 + \ldots\ .
\eea
Here, $H_k$ is the harmonic number $H_k := \sum_{n=1}^k \frac{1}{n}$ and $H_0 := 0$.  We find that the K\"ahler transformation parameter agrees (numerically, to high order) with the fundamental period computed by \rodland{} \cite{Rodland9801}, which we present here for convenience\footnote{Note that this suggests a nontrivial combinatorial identity between the sums in (\ref{eqn:OurRodPeriod}) and (\ref{eqn:RodPeriod}).}
\bea
\label{eqn:RodPeriod}
X^0_{\textrm{\rodland}}(z) &=& \sum_{m_i \in \mathbb{N}_0^4} (-1)^{m_2+m_4} {m\choose m_2} {m\choose m_4} {m+m_3 \choose m} {m+m_2+m_3 \choose m_1 ,\, m_2+m_3, \, m_2+m_3+m_4}  \nonumber \\
&& \qquad \qquad \qquad \times { m+m_3+m_4 \choose m_1,\, m_3+m_4, \, m_2+m_3+m_4} \, z^m \, \bigg|_{m=\sum_i m_i} \ , \\
& = & 1 + 5\,z + 109\, z^2 + 3\,317\, z^3 + 121\,501\, z^4 + 4\,954\,505\, z^5 + \ldots \ .
\eea
Following the procedure described in Section \ref{sec:procedure}, we find Gromov--Witten invariants that agree with the values in the literature \cite{batyrev1998conifold}.

\noindent \underline{\bf Pfaffian Phase: $r \ll 0$}

In this phase, we must close the contours in the upper-half $\sigma_0$ and $\sigma_1$ planes.  There is a slight subtlety, however, since after performing the $\sigma_0$ integral by the method of residues, the $\sigma_1$ integrand will lack a convergence factor, causing a slight power-law divergence in the integrand.  Since we are closing the $\sigma_1$ contour in the upper half plane, we can regulate this by including a convergence factor $e^{i\delta\sigma_1}$, then taking the limit as $\delta\rightarrow 0^+$ at the end.  This yields
\bea
Z &=& \frac{(z\bar{z})^{2\rr-1}}{2} \lim_{\delta\rightarrow 0^+}    \oint \frac{d^2\epsilon}{(2\pi i)^2} \
\bigg(\frac{\pi^7\sin^7\pi(\epsilon_0+\epsilon_1)}{\sin^7\pi\epsilon_0\sin^7\pi\epsilon_1}\bigg) \ (z\bar{z})^{-\epsilon_0} \nonumber \\
&& \quad \times  \left| \sum_{K,k\geq 0} (-e^{-\delta})^{k}  (-z)^{-K}\big( 1+K+2k+\epsilon_0+2\epsilon_1 \big)\,    \frac{\Gamma(1+K+k+\epsilon_0+\epsilon_1)^7}{\Gamma(1+K+\epsilon_0)^7 \, \Gamma(1+k+\epsilon_1)^7} \right|^2  \, ,  \qquad\quad
\eea
where the complex conjugation does not act on $\epsilon_{0,1}$.  Notice that the sum over $k$ for fixed $K$ is now infinite, rather than finite.  Also, note that the sum over $k$ is at its radius of convergence when $\delta=0$, but the alternating sign $(-1)^k$ ensures that the final result is insensitive to how we take the $\delta\rightarrow 0$ limit, so we can safely remove the convergence factor.

Unfortunately, the fact that the sums over $k$ are infinite makes the calculation of Gromov--Witten invariants computationally intensive.  A simple computation demonstrates that the coefficient of $\log^3\bar{z}$, corresponding to the K\"ahler transformation, is
\be
X^0_{\textrm{Pfaff}}(z) = z^{-1} + 17\,z^{-2} + 1\,549\,z^{-3} + 215\,585\,z^{-4} + 36\,505\,501\, z^{-5} + 6\,921\,832\,517\, z^{-6}+ \ldots \ .
\ee
Indeed, this is annihilated to sixth order by the degree 5 Picard--Fuchs operator conjectured by \rodland.  Combined with the agreement of Gromov--Witten invariants in the Grassmann phase, we consider this to be strong evidence that the conjecture holds in the Pfaffian phase as well.

\section{The determinantal Gulliksen-Neg\aa{}rd Calabi--Yau threefold}
\label{sec:GN}

\begin{figure}[t]
\centering
\includegraphics[width=1.75in]{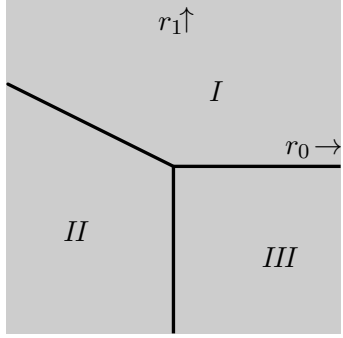}
\put(-50,88){\emph{I}}
\put(-105,35){\emph{II}}
\put(-30,25){\emph{III}}
\put(-21,68){$r_0\! \rightarrow$}
\put(-69,115){$r_1$}
\put(-61,116){$\uparrow$}
\caption{The \emph{classical} GLSM moduli space, as a function of the FI parameters $(r_0,r_1)$, of the GLSM that describes the GN Calabi--Yau threefold.  The grey shading represents regions of the FI parameter space where we have a classical geometric description. The solid black lines indicate boundaries of the respective geometric phases, along which classical Coulomb branches emerge. The \emph{I}-\emph{II} phase boundary has slope $-\frac{1}{2}$.}
\label{fig:GN-phases}
\end{figure}

We now consider a determinantal Calabi--Yau threefold $Y$ that is defined as the rank two locus of a generic section of $\rm{Hom}(\cO_{\P^7}^{\oplus 4}, \cO_{\P^7}(1)^{\oplus 4})$ --- such a section is described by a $4\times 4$ matrix $A(\phi) = \sum_a A^a \phi_a $ of linear forms in the homogeneous coordinates $\phi_{a=1,\cdots,8}$ of $\P^7$, where the $A^a$ are eight constant $4\times 4$ matrices. The ideal describing the variety is generated by the $3\times 3$ minors of the defining matrix $A(\phi)$.  The resolution of such an ideal was first studied by Gulliksen and Neg\aa{}rd \cite{gulliksen1971}, so we will refer to this determinantal variety $Y$ as the GN Calabi--Yau threefold.
A GLSM construction for determinantal varieties has recently been proposed by the present authors \cite{Jockers:2012zr}, so we refer the reader to that paper for further details. Following the notation of that paper, we will use the ``PAX'' model to describe the GN Calabi--Yau threefold, with gauge group, matter content, and R-charges summarized in Table~\ref{tab:GN}.
\begin{table}[t]
\begin{tabular}{|c||c|c|c|}
\hline
Field 	& $U(1)$ & $U(2)$   			& $U(1)_{\textsc{v}}$  \\
\hline
\hline
$\Phi_a$ & $+1$	&	$1_0$				& $2\rr_\phi$		 \\
$P_i$		& $-1$	& $\Box_{+1}$ & $2-2\rr_x-2\rr_\phi$		\\
$X_i$		& $0$		& $\overline{\Box}_{-1}$				& $2\rr_x$		 \\
\hline
\end{tabular}
\centering
\caption{Gauge group, matter content and R-charges of a GLSM describing the GN Calabi--Yau threefold in $\P^7$. Note that $a=1,\ldots,8$, and $i=1,\ldots,4$.}
\label{tab:GN}
\end{table}
The GLSM has a superpotential
\begin{equation}
W = \tr\big(P A(\Phi) X\big)\, ,
\end{equation}
where the field $P$ is a $2 \times 4$ matrix whose columns, $P_{i}$, transform in the fundamental of $U(2)$, and $X$ is a $4 \times 2$ matrix whose rows, $X_i$, transform in the anti-fundamental representation of $U(2)$, where $i=1,\ldots,4$.  The classical vacuum moduli space is defined by the following D-term equations
\begin{eqnarray}
U(1): &  \sum_a |\phi_a|^2 - \tr(p^\dag p) = r_0 \, , \\
U(2): & p p^\dag - x^\dag x  = r_1 \mathds{1}_{2\times 2} \, ,
\label{eq:Uk-D}
\end{eqnarray}
and F-term equations
\begin{eqnarray}
(A^a\phi_a) x = 0\, , \qquad p \, (A^a\phi_a) = 0\, , \qquad \tr (pA^ax) = 0\, . \label{eq:Uk-F}
\end{eqnarray}
The classical solutions of these equations can be divided into phases depending on the values of the FI parameters and is depicted in Figure \ref{fig:GN-phases} --- quantum mechanically, the phase structure is corrected.  The three asymptotic regions, labelled \emph{I}, \emph{II}, \emph{III}, correspond to distinct geometric phases.\footnote{Note that this corrects an oversight in an earlier version of this paper where the slope of the \emph{I}-\emph{II} phase boundary was stated to be $-1$.  We thank K. Hori for pointing this out to us.} The phase $r_0+2r_1 \gg 0$ and $r_1 \gg 0$, for example, is associated with the incidence correspondence
\begin{equation} \label{GNCY}
  Y \,=\, \big\{ \, (\phi,p) \in \P^7 \times G(2,4)\ \big| \  p\,  (A^a\phi_a) = 0\, \big\}\ .
\end{equation}

Before embarking on the calculation of the partition function, we review some facts about the geometry of the GN Calabi--Yau threefold $Y$ from \cite{gross2001calabi,Bertin0701,Kapustka0802,Jockers:2012zr}.  The Hodge numbers $h^{1,1}=2$ and $h^{2,1}=34$ are calculated with the aid of \cite{MR1487235}.  There are two maps from $Y$ to the Grassmannian $G(2,4)$:  in one case, we send $\phi$ to the kernel of the matrix $A(\phi)$, and in the other case we send it to the kernel of the transposed matrix $A(\phi)^T$.  A ``Schubert'' divisor $\divisorsigma$, which generates $H^2(G(2,4),\mathbb{Z})$, therefore gives rise to two different divisors $\divisorsigma_0$ and $\divisorsigma_1$ on $Y$, depending on which map to $G(2,4)$ we use.  There is a third divisor on $Y$ coming from the hyperplane class $H$ on $\mathbb{P}^7$; since $h^{1,1}=2$, there must be a relation among these, and it can be shown to be $\divisorsigma_0+\divisorsigma_1=2H$.

The image of $Y$ in the Grassmannian $G(2,4)$ under either of the maps is a threefold with $56$ ordinary double points (i.e., conifold points).  This shows that $\divisorsigma_0$ and $\divisorsigma_1$ each lie on the boundary of the K\"ahler cone; since the K\"ahler cone is two-dimensional, they must generate its two boundary edges.  In fact, as shown in \cite{Kapustka0802}, the curves $C_{\mu=0,1}$, which are contracted to nodes by $\divisorsigma_\mu$, are lines in $\mathbb{P}^7$.  Thus, we have $H\cdot C_\mu=1$ and $\divisorsigma_\mu\cdot C_\mu=0$, which easily imply (using $\divisorsigma_0+\divisorsigma_1=2H$) that $\divisorsigma_{\mu}\cdot C_{\nu} = 2-2\delta_{\mu\nu}$.

As we shall see later, there is another type of curve $\Gamma$ on $Y$ which is a line in $\mathbb{P}^7$, satisfying $\divisorsigma_\mu\cdot\Gamma=H\cdot\Gamma=1$.  (Note the homology relation $2\Gamma = C_0 + C_1$.)
The existence of $C_0$, $C_1$, and $\Gamma$, shows that any
$\mathbb{Q}$-linear combination of $H$ and $\divisorsigma_1$ that has
integral intersection numbers with all curves on $Y$ must be a
$\mathbb{Z}$-linear combination (since $C_1$ and $\Gamma-C_1$ form
a dual pair to $H$ and $\divisorsigma_1$ under the intersection pairing).  Thus, $\{H,\divisorsigma_1\}$ is a basis
for $H^2(Y,\mathbb{Z})$, as asserted in \cite{Jockers:2012zr}.

The triple intersection numbers of this basis of divisors are also
easy to calculate \cite{Jockers:2012zr}, yielding
\begin{equation}  \label{eq:Inter1}
   H^3 = 20\ , \qquad H^2\cdot \divisorsigma_1 = 20\ , \qquad H\cdot\divisorsigma_1^2 = 16 \ , \qquad \divisorsigma_1^3 = 8 \ .
\end{equation}
This result can also be expressed in terms of the K\"ahler cone generators,
$\divisorsigma_0$ and $\divisorsigma_1$, as
\begin{equation} \label{eq:Inter2}
  \divisorsigma_0^3 = 8 \ , \qquad \divisorsigma_0^2\cdot\divisorsigma_1 = 24\ , \qquad \divisorsigma_0\divisorsigma_1^2 = 24\ , \qquad \divisorsigma_1^3=8 \ .
\end{equation}

\subsection{Partition function and instanton numbers}

The ``PAX'' GLSM for the GN Calabi--Yau threefold has two FI parameters, $r_0$ and $r_1$, that give rise to the two K\"ahler parameters of the GN Calabi--Yau variety. In terms of these K\"ahler parameters, the partition function is given by the formula
\begin{equation}
\begin{aligned} \label{eqn:Z-GN}
Z_{\textrm{GN}} =  \frac{1}{2}  \sum_{m_0,m_1,m_2 \in \Z} \int \frac{d\sigma_0 d\sigma_1 d\sigma_2}{(2\pi)^3}\,&  \left[ \frac{(m_1-m_2)^2}{4}+(\sigma_1-\sigma_2)^2 \right]\, e^{-4\pi i r_0 \sig_0 - i\theta_0 m_0} \\
& \quad \times e^{-4\pi i r_1(\sig_1+\sig_2) -i\theta_1(m_1+m_2)}\ (Z_X)^4\, (Z_P)^4\, (Z_\Phi)^8\ ,
\end{aligned}
\end{equation}
where
\begin{equation}
\begin{aligned}
Z_X & = \frac{\Gamma\big(\rr_x+i\sig_1+\tfrac{m_1}{2}\big)}{\Gamma\big(1-\rr_x - i\sig_1+\tfrac{m_1}{2}\big)}\, \frac{\Gamma\big(\rr_x+i\sig_2+\tfrac{m_2}{2}\big)}{\Gamma\big(1-\rr_x-i\sig_2+\tfrac{m_2}{2}\big)}\  ,  \\
Z_\Phi & = \frac{\Gamma\big(\rr_\phi-i\sig_0-\tfrac{m_0}{2}\big)}{\Gamma\big(1-\rr_\phi+i\sig_0-\tfrac{m_0}{2}\big)}\ , \\
Z_P & = \frac{\Gamma\big(1-\rr_x-\rr_\phi +i(\sig_0-\sig_1)+\tfrac{m_0-m_1}{2}\big)}{\Gamma\big(\rr_x+\rr_\phi-i(\sig_0-\sig_1)+\tfrac{m_0-m_1}{2}\big)} \, \frac{\Gamma\big(1-\rr_x-\rr_\phi +i(\sig_0-\sig_2)+\tfrac{m_0-m_2}{2}\big)}{\Gamma\big(\rr_x+\rr_\phi-i(\sig_0-\sig_2)+\tfrac{m_0-m_2}{2}\big)}\  .
\end{aligned}
\end{equation}
The partition function can be evaluated analogously to the quintic example.  In phase $\mathit{I}$, expanded around $r_0 + 2r_1 \gg 0$ and $r_1 \gg 0$, the result is
\begin{equation} \label{ZGN}
\begin{aligned}
Z_{\textrm{GN}} &= -\frac{1}{2} \oint \frac{d\epsilon_0\,d\epsilon_1\,d\epsilon_2}{(2\pi i)^3} \ \frac{\pi^8 \sin^4(\pi \epsilon_0+\pi \epsilon_1)\, \sin^4(\pi \epsilon_0+\pi \epsilon_2) }{\sin^8 \pi\epsilon_0 \, \sin^4 \pi \epsilon_1 \, \sin^4 \pi \epsilon_2} \ (z \zb)^{\epsilon_0} (w \wb)^{\epsilon_1+\epsilon_2}  \\
& \qquad \Bigg\vert \sum_{K_0,K_1=0}^\infty z^{K_0}\, w^{K_1} \sum_{k_1=0}^{K_1}  (2k_1-K_1+\epsilon_1-\epsilon_2)  \, \frac{\Gamma(1+K_0+k_1+\epsilon_0+\epsilon_1)^4}{\Gamma(1+K_0+\epsilon_0)^4\,\Gamma(1+k_1+\epsilon_1)^4 }   \\
&  \qquad \frac{ \Gamma(1+K_0+K_1-k_1+\epsilon_0+\epsilon_2)^4 }{\Gamma(1+K_0+\epsilon_0)^4\,\Gamma(1+K_1-k_1+\epsilon_2)^4 } \Bigg\vert^2  \ ,
\end{aligned}
\end{equation}
where complex conjugation does not act on $\epsilon_{0,1,2}$, and where we have defined the algebraic coordinates
\begin{equation}
  z \,=\, e^{(-2\pi r_0 + i \theta_0)+2(-2\pi r_1 + i \theta_1)} \ , \qquad
  w\,=\, e^{-2\pi r_1 + i \theta_1} \ .
\end{equation}

From the partition function \eqref{ZGN}, we extract the K\"ahler transformtion $X^0$ and the flat IR coordinates $t^\ell$, $\ell=0,1$, given in terms of the algebraic coordinates $z$ and $w$ by
\begin{equation}
\begin{aligned}
\label{eqn:GN-fund}
     X^0(z,w)\,&=\,1 + 2 w + z + 3 w^2 + z^2 + 4 w^3 - 14 w^2 z - 54 w z^2 + z^3 +  \ldots \ , \\
     2\pi i\,t^0(z,w) \,&=\,  \log z + 4 w + 2 w^2 - 20 w z + \frac43w^3 - 72 w^2 z - 92 w z^2 + \ldots   \ , \\
     2\pi i\,t^1(z,w) \,&=\,  \log (-w) + 4 z + 16 w z + 2 z^2 + 36 w^2 z - 128 w z^2 + \tfrac43 z^3 + \ldots  \ .
\end{aligned}
\end{equation}
This allows us to expand the partition function $Z_{\textrm{GN}}$ in flat coordinates $t^\ell$ as
\begin{equation} \label{ZGNflat}
\begin{aligned}
  Z_{\textrm{GN}} =& -\tfrac{20i}{3!}(t^0-\bar t^0)^3-\tfrac{8i}{3!} (t^1-\bar t^1)^3
  -\tfrac{20i}2 (t^0-\bar t^0)^2(t^1-\bar t^1)-\tfrac{16i}{2} (t^0-\bar t^0)(t^1-\bar t^1)^2-\tfrac{64\,\zeta(3)}{4\pi ^3} \\
  &\quad + \tfrac{2 i}{(2\pi i)^3}\Big( 56\,(q_0+\bar q_0) + 192\,(q_0 q_1+\bar q_0 \bar q_1) + \tfrac{56}{2^3}\,(q_0^2+\bar q_0^2) + \ldots  \Big) \\
  &\quad\quad -\tfrac{i}{(2\pi i)^2}\Big( 56\,(t_0-\bar t_0) (q_0+\bar q_0) + 192\,(t^0+t^1-\bar t^0-\bar t^1) (q_0 q_1+\bar q_0 \bar q_1)+\ldots \Big)  \ ,
\end{aligned}
\end{equation}
where $q_\ell=e^{2\pi i t^\ell}$. Here, the partition function is canonically normalized with respect to the Euler characteristic $\chi(Y) = -64$.

From the classical part of the partition function \eqref{ZGNflat}, we readily extract the intersection numbers \eqref{eq:Inter1} of the integral generators $\omega_\ell$, $\ell=0,1$, of the second cohomology. Thus, we find agreement with the intersection numbers calculated in \cite{Jockers:2012zr} by identifying $\omega_0$ and $\omega_1$ with the hyperplane class $H$ of $\P^7$ and the Schubert class $\divisorsigma_1$ of $G(2,4)$ in the incidence correspondence \eqref{GNCY}, respectively. This is precisely the identification we expect since the ``PAX'' model relates directly to the geometry of the embedding into $\mathbb{P}^7$, which corresponds to $H$, and of the embedding into $G(2,4)$ corresponding to $\divisorsigma_1$.  This furnishes a nontrivial check on the classical part of the partition function~\eqref{ZGNflat}.

\def\bstrut{\rule[-1.2ex]{0ex}{3.5ex}}
\begin{table}[t]
\begin{center}
\valign{\vfil#\vfil\cr
\small{
\vbox{\offinterlineskip
\halign{\strut\vrule~\hfil$#$~\vrule\vrule&\hfil$\,#$\,&\hfil\quad$#$\,&\hfil\quad$#$\,&\hfil\quad$#$\,&\hfil\quad$#$\,&\hfil\quad$#$\,&\hfil\quad$#$\ \ \vrule\cr
\noalign{\hrule}
\widetilde{N}_{m_0,m_1} & m_0\!=\!0 & \slfrac12 & 1 & \slfrac32 & 2 & \slfrac52 & 3\cr
\noalign{\hrule\hrule}
m_1\!=\!0 & - && 56 && 0 && 0 \cr
\slfrac12 && 192 && 896 && 192 & \cr
1 &  56 && 2\,544 && 23\,016 &&  41\,056 \cr
\slfrac32 && 896 && 52\,928 && 813\,568 &  \cr
2 & 0 && 23\,016 && 1\,680\,576 && 35\,857\,016 \cr
\slfrac52 && 192 && 813\,568 && 66\,781\,440 & \cr
3 & 0 && 41\,056 && 35\,857\,016 && 3\,074\,369\,392 \cr
\slfrac72 && 0 && 3\,814\,144 &&  1\,784\,024\,064 & \cr
4 & 0 && 23\,016 && 284\,749\,056 && 96\,591\,652\,016 \cr
\slfrac92 && 0 && 6\,292\,096 &&20\,090\,433\,088 & \cr
5 & 0 && 2\,544 &&933\,789\,504 && 1\,403\,214\,088\,320 \cr
\slfrac{11}2 && 0 && 3\,814\,144 &&105\,588\,804\,096 &\cr
6 & 0 && 56 && 1\,371\,704\,192 &&10\,388\,138\,826\,968 \cr
\slfrac{13}2 && 0 && 813\,568 &&277\,465\,693\,248&\cr
7 & 0 && 0 && 933\,789\,504 &&41\,598\,991\,761\,344\cr
\slfrac{15}2 && 0 && 52\,928 && 380\,930\,182\,784&\cr
8 & 0 && 0 && 284\,749\,056 &&93\,976\,769\,192\,864\cr
\slfrac{17}2 && 0 && 896 && 277\,465\,693\,248&\cr
9 & 0 && 0 &&  35\,857\,016&&122\,940\,973\,764\,384\cr
\slfrac{19}2 && 0 && 0 &&105\,588\,804\,096&\cr
10 & 0 && 0 &&  1\,680\,576&&93\,976\,769\,192\,864\cr
\slfrac{21}2 && 0 && 0 &&20\,090\,433\,088&\cr
11 & 0 && 0 &&  23\,016&&41\,598\,991\,761\,344\cr
\slfrac{23}2 && 0 && 0 &&1\,784\,024\,064&\cr
12 & 0 && 0 && 0&&10\,388\,138\,826\,968\cr
\slfrac{25}2 && 0 && 0 &&66\,781\,440& \cr
13 & 0 && 0 && 0&&1\,403\,214\,088\,320\cr
\slfrac{27}2 && 0 && 0 &&813\,568&\cr
14 & 0 && 0 && 0&&96\,591\,652\,016\cr
\slfrac{29}2 && 0 && 0 &&192&\cr
15 & 0 && 0 && 0&&3\,074\,369\,392\cr
\slfrac{31}2 && 0 && 0 &&0&\cr
\noalign{\hrule}}
}}\cr}
\centering
\caption{Predictions for genus zero Gromov--Witten invariants of the determinantal GN Calabi--Yau threefold $Y$.}
\label{tab:GNGWbis}
\end{center}
\end{table}

From the non-perturbative terms of the partition function \eqref{ZGN}, we extract the genus zero worldsheet instanton corrections $N_{\ell_0,\ell_1}$ (organized by instanton number with respect to the basis $\{H,\divisorsigma_1\}$ of second cohomology), which count the number of curves
in the homology class $\ell_0 C_1 + \ell_1 (\Gamma-C_1)$.  The structure of these numbers is more apparent if we use a basis of the Mori cone instead.  We can rewrite the homology class as
\begin{equation}
  \ell_0 C_1 + \ell_1 (\Gamma-C_1) =
  \ell_0 C_1 + \ell_1 \left(\tfrac12(C_0+C_1)-C_1\right)
  = \left(\ell_0-\tfrac12\ell_1\right)C_1+\tfrac12\ell_1 C_0 \ .
\end{equation}
Thus, we can label the new homology classes $m_0C_1+m_1C_0$ with $m_0$
and $m_1$ nonnegative half-integers whose sum $m_0+m_1$ is an integer.

We call the resulting instanton correction $\tilde{N}_{m_0,m_1}$, and
we have listed the values for these in Table~\ref{tab:GNGWbis}.
We conjecture that the quoted integers $\tilde{N}_{m_0,m_1}$ are the integral genus zero Gromov--Witten invariants of the threefold $Y$ in the homology class $m_0\, [C_1] + m_1\, [C_0] \in H_2(Y,\Z)$.\footnote{As proposed in ref.~\cite{Kim:06}, an appropriate version of the quantum hyperplane theorem 
\cite{alg-geom/9612001,arXiv:alg-geom/9712008} (also known as the quantum 
restriction formula \cite{summing}) should allow one to calculate the genus zero Gromov--Witten invariants of 
any Calabi--Yau threefold that arises as the zero locus of a vector bundle over a geometric quotient $V//G_{\mathbb{C}}$ (where a complex Lie group $G_{\mathbb{C}}$ acts on the vector space $V$) by using the conjectural J-function of $V//G_{\mathbb{C}}$, which is also proposed in \cite{Kim:06}. Since the GN Calabi--Yau threefold falls into this class of geometries \cite{Jockers:2012zr}, it would be interesting to compare our predictions of Gromov--Witten invariants with this proposal.}   

\subsection{Gromov--Witten invariants}
\label{sec:GWinv}

Our first observation is a prediction of $\tilde{N}_{1,0}=56$ for the number of lines in $\P^7$ contracted by one of the maps to $G(2,4)$, and $\tilde{N}_{0,1}=56$ for the number of lines contracted by the other map.
As discussed in detail in~\cite{Jockers:2012zr}, this is precisely the number
of contracted lines predicted by the geometry.  In the first case, these $56$ lines give rise to 56 nodal points in the singular Calabi--Yau threefold
\begin{equation}
\label{eqn:Ysing}
 Y_{\textrm{sing}}\,=\, \{ \ p \in G(2,4) \ | \ \det \cA = 0  \ \} \ , \qquad \left(\cA \right)^a_{i\alpha} \equiv A^a_{ij}p_{\alpha i} \ ,
\end{equation}
by blowing down the curve class dual to $\divisorsigma_1$; the second case is similar.

Curves $\Gamma$ on $Y$ that satisfy $\divisorsigma_0\cdot \Gamma
=\divisorsigma_1\cdot\Gamma = H\cdot\Gamma=1$ are constructed as follows.
First, since $H\cdot\Gamma=1$, the curve must be a line
in $\mathbb{P}^7$.
Moreover, each projection of the curve to $G(2,4)$ must be a Schubert
cycle since its intersection number with the Schubert divisor is
$1$.  Let us fix a four dimensional space $\mathbb{C}^4$.
Those Schubert cycles are defined geometrically by choosing
a hyperplane $V\subset\mathbb{C}^4$ and then choosing
a line $W\subset V$; the Schubert cycle consists of all $2$-planes
$\Pi$ such that $W\subset \Pi \subset V$.
The dual space $(\mathbb{C}^4)^*$ must also have similar data,
specified by a dual hyperplane $\tilde{W}\subset (\mathbb{C}^4)^*$
and a dual line $\tilde{V}\subset \tilde{W}$.

Given a line $\mathbb{L}\subset \mathbb{P}^7$ and data $\{V, W, \tilde{V},\tilde{W}\}$
specifying Schubert $1$-cycles on the dual Grassmannians, we can choose
coordinates on $\mathbb{C}^4$ and $(\mathbb{C}^4)^*$ such that $V$ is 
spanned by $e_1$, $W$ is spanned by $e_1$, $e_2$, $e_3$, $\tilde{V}$
is spanned by $\tilde{e}_1$, and $\tilde{W}$ is spanned by 
$\tilde{e}_1$, $\tilde{e}_2$, $\tilde{e}_3$.  In order for  a $4\times4$ matrix $A$
of linear forms on $\mathbb{P}^7$ to contain the line $\mathbb{L}$ in 
such a way that $\mathbb{L}$ projects to the given Schubert cycles
in the two Grassmannians, we need for $A$ to restrict on $\mathbb{L}$
 to a matrix of the form
\begin{equation} \label{eq:L192}
 A|_{\mathbb{L}}=\begin{bmatrix} 0&0&0&0\\ 0&0&0&\ell_1(u,v)\\
0&0&0&\ell_2(u,v)\\
0&\ell_3(u,v)&\ell_4(u,v)&\ell_5(u,v)
\end{bmatrix} \ ,
\end{equation}
where the $\ell_\mu(u,v)$ are linear forms on $\mathbb{L}$ such that 
$\{\ell_1, \ell_2\}$ and $\{\ell_3,\ell_4\}$ are linearly independent sets.

Note that the kernel of $A$ at the point $[u,v]\in \mathbb{L}$ is spanned by
$e_1$ and $\ell_2(u,v)e_2-\ell_1(u,v)e_3$, so it lies in the specified
Schubert
$1$-cycle.
Similarly, the kernel of $A^T$ at $[u,v]$ is spanned by
$\tilde{e}_1$ and $\ell_4(u,v)\tilde{e}_2-\ell_3(u,v)\tilde{e}_3$.

The prediction of our calculation is that if $Y$ is fixed and generic, 
there are exactly $192$ such lines on $Y$. We confirm this number of
lines explicitly in Appendix~\ref{sec:lines}.

\subsection{The extremal transition}

Instead of desingularizing the singular threefold $Y_{\textrm{sing}}$ (\ref{eqn:Ysing}) with a small resolution to recover the Calabi--Yau threefold $Y$, we can deform the complex structure to obtain a smooth deformed Calabi--Yau threefold $\widehat Y$. Such a transition from a resolved Calabi--Yau threefold to a deformed Calabi--Yau threefold has been studied thoroughly in the literature and is called an extremal transition \cite{clemens1983double,MR848512,MR909231,Candelas:1989ug,Greene:1995hu}. 

Here the deformed Calabi--Yau threefold $\widehat Y$ is identified with the complete intersection $\P^5[2,4]$. To see this, realize the hypersurface equation $\det \cA(p)=0$ of $Y_{\textrm{sing}}$ in $G(2,4)$ as a polynomial of degree four in terms of the Pl\"ucker coordinates $[u_{12}:u_{13}:u_{14}:u_{23}:u_{24}:u_{34}]$, with $u_{ij}=p_{1i}p_{2j}-p_{2i}p_{1j}$ furnishing homogeneous coordinates of $\P^5$ that satisfy quadratic Pl\"ucker relation $u_{14}u_{23}-u_{13}u_{24}+u_{12}u_{34}=0$ along the $G(2,4)$ hypersurface.  Thus, the hypersurface equation $\det \cA(u)=0$ --- which is a non-generic homogeneous polynomial of degree four --- together with the Pl\"ucker relation give rise to the singular complete intersection Calabi--Yau threefold $Y_{\textrm{sing}}$ in $\P^5$.  Deforming these two equations generically yields a smooth complete intersection in $\P^5[2,4]$, which we identify as the deformed Calabi--Yau threefold $\widehat Y$ with $h^{1,1}(\widehat Y)=1$ and $h^{2,1}(\widehat Y)=89$.

Along the described extremal transition from the threefold $Y$ to the threefold $\widehat Y$, the homology class $[C_1] \in H_2(Y,\Z)$ vanishes so that $H_2(\widehat Y,\Z)$ can only be generated by the two cycle $[C_0]$ dual to the Schubert class $\divisorsigma_1$ of $G(2,4)$. As a consequence, the grading of the Gromov--Witten invariants $\tilde{N}_{m_0,m_1}$ with respect to the degree $m_0$ of the homology cycle $[C_1]$ disappears and the genus zero instanton invariants obey \cite{MR1839289}
\begin{equation}
  N_{\ell}(\widehat Y)\,=\, \sum_{\substack{m=0\\ m+\ell \text{ even}}}^{\infty} \tilde{N}_{m/2,\ell/2} \ .
\end{equation}
Thus, we extract from the first few numbers $\tilde{N}_{m_0,m_1}$, via the discussed extremal transition, the integral Gromov--Witten invariants 
\begin{equation}
   N_\ell(\widehat Y)=1\,280\, ,\ \ 92\,288 \, ,\ \ 15\,655\,168 \, ,\ \ 3\,883\,902\,528 \, ,\ \ 1\,190\,923\,282\,176\, , \ \ldots \ .
\end{equation}
These Gromov--Witten invariants $N_{\ell}(\widehat Y)$ of $\P^5[2,4]$ are in agreement with~\cite{MR1201748}, where these numbers have been computed independently and by different means, serving as a nontrivial consistency check on the conjectured Gromov--Witten invariants $\tilde{N}_{m_0,m_1}$.

\subsection{Symmetries of the Gromov--Witten invariants}

A quick glance at Table~\ref{tab:GNGWbis} shows two 
symmetries
among our predicted Gromov--Witten invariants:  we have
$\tilde{N}_{m_0,m_1}=\tilde{N}_{m_1,m_0}$ for all $(m_0,m_1)$, and
 $\tilde{N}_{m_0,m_1}=
\tilde{N}_{m_0,6m_0-m_1}$ whenever $(m_0,m_1)\ne(0,1)$.  
(There is also a third symmetry which is a consequence of the other two:
$\tilde{N}_{m_0,m_1}=\tilde{N}_{6m_1-m_0,m_1}$ whenever $(m_0,m_1)\ne(1,0)$.)

The first of these is easy to explain.  If we map the defining matrix $A$
of a GN Calabi--Yau threefold
to its transpose $A^T$, we get another GN Calabi--Yau threefold.  That is,
taking the transpose of the matrix $A$ defines an automorphism of order two
on the complex structure moduli space.  This automorphism acts nontrivially
on $H^2(Y)$, exchanging the K\"ahler cone generators $\divisorsigma_0$ and
$\divisorsigma_1$.  Since the Gromov--Witten invariants must be the same after
acting by this automorphism, we see that $\tilde{N}_{m_1,m_0}=\tilde{N}_{m_0,m_1}$.

The second symmetry is due to the existence of a flop $Y^+$ of $Y$ along
the curves in class $[C_0]$ which is diffeomorphic to the original Calabi--Yau
variety (this was noted in \cite{Jockers:2012zr}).\footnote{Other examples of this phenomenon can be seen in \cite{2param1,Hosono:2011vd}.}
As explained in \cite{beyond}, this occurs due to the following:

The K\"ahler cone of $Y$ is generated by $\divisorsigma_0$ and $\divisorsigma_1$; if
we write the three-point correlation functions with respect to 
coordinates  $u_0$ and $u_1$, which are adapted to $\divisorsigma_0$ and $\divisorsigma_1$,
we get expressions of the form
\begin{equation} \label{eq:3point}
 \langle \alpha,\beta,\gamma\rangle 
= \alpha \cdot \beta\cdot \gamma 
+ \sum_{C=m_0C_1+m_1C_0} \frac{u_0^{m_0}u_1^{m_1}}{1-u_0^{m_0}u_1^{m_1}}
(\alpha\cdot C)(\beta\cdot C)(\gamma\cdot C)
\tilde{N}_{m_0,m_1} \ .
\end{equation}
As shown in \cite{beyond}, if we remove the term $C=C_0$ from the summation,
then the resulting expression
\begin{equation}\label{eq:3pointsum}
 \sum_{\substack{C=m_0C_1+m_1C_0\\(m_0,m_1)\ne (0,1)}} \frac{u_0^{m_0}u_1^{m_1}}{1-u_0^{m_0}u_1^{m_1}}
(\alpha\cdot C)(\beta\cdot C)(\gamma\cdot C)
\tilde{N}_{m_0,m_1}.
\end{equation}
must be valid throughout  the union of the K\"ahler cones of $Y$ and of $Y^+$.
Note that the flop does not affect rational curves in classes other than $[C_0]$, so the only Gromov--Witten invariant which changes under this flop is $\tilde{N}_{0,1}$.

The K\"ahler cone of $Y^+$ is generated by $\divisorsigma_1$
and some other vector $\divisorsigma_2$ which can be written in the form
$\divisorsigma_2=\lambda\divisorsigma_1- \divisorsigma_0$.  
Since $\divisorsigma_1 \cdot (m_0C_1+c_1C_0)=m_0$
and $(\lambda\divisorsigma_1-\divisorsigma_0) \cdot (m_0C_1+c_1C_0)=\lambda m_0-m_1$,
and since \eqref{eq:3pointsum} must be well-defined throughout the union
of the K\"ahler cones,
the Gromov--Witten invariants $\tilde{N}_{m_0,m_1}$ can only be nonzero
if $(m_0,m_1)= (0,1)$ or $\lambda m_0-m_1\ge0$.  That is the general
feature of the instanton sums near a flop.

However, if the flopped variety $Y^+$ is isomorphic to $Y$, then there is
a restriction not only on
the set of nonzero Gromov--Witten invariants but also on their values:  we
must have $\tilde{N}_{m_0,m_1}=\tilde{N}_{m_0,\lambda m_0-m_1}$ unless
$(m_0,m_1)=(0,1)$ (in order to get an identical sum in \eqref{eq:3pointsum}
after the flop).  This is precisely the symmetry we observe for
the predicted invariants (with 
$\lambda=6$), so we see another important compatibility property between
the predicted invariants and the geometry.

\subsection{Picard--Fuchs system and singular loci}

Using the fundamental period (\ref{eqn:GN-fund}), to high order in $z$ and $w$, we find two Picard--Fuchs operators:
\be
\mathcal{L}_1 = \sum_{j=0}^2 \sum_{k=0}^{2-j} p_{jk}(z,w) \Theta_z^j \Theta_w^k \ ,  \qquad  \mathcal{L}_2 = \sum_{j=0}^3 \sum_{k=0}^{3-j} r_{jk}(z,w) \Theta_z^j \Theta_w^k \ ,
\ee
where $\Theta_z$ and $\Theta_w$ are logarithmic derivatives, and the polynomials are
\begin{equation}
\begin{aligned}
p_{00} &= -5 w+20 w^2-25 w^3+10 w^4-3 z+43 w z-41 w^2 z+3 z^2\ ,   \\
p_{01} &= -10 w+30 w^2-30 w^3+10 w^4-7 z+52 w z-45 w^2 z+5 z^2\ , \\
p_{10} &= 6 (5 w^2 - 10 w^3 + 5 w^4 - z + 16 w z - 16 w^2 z + z^2)\ ,  \\
p_{02} &= \tfrac{1}{2} \left( 1-2 w+w^2-5 z\right) \left( 5-10 w+5 w^2-z\right)\ , \\
p_{11} &= -5+30 w^2-40 w^3+15 w^4+52 w z-52 w^2 z+5 z^2 \ , \\
p_{20} &= 3+8 w-2 w^2-32 w^3+23 w^4-6 z+56 w z-58 w^2 z+3 z^2\ ,
\end{aligned}
\end{equation}
and
\be
\begin{array}{lcl}
r_{00} =   5 w (-1+4 w) \, ,  &  \qquad\qquad  &  r_{01} =  -15 w+30 w^2-z \, , \\
r_{10} =   5 w (-1+14 w) \, , & \qquad\qquad &  r_{02} =  \frac{3}{2} \left(-10 w+10 w^2-z\right)\, , \\
r_{11} =  2 \left(-10 w+35 w^2-z\right)  \, ,  & \qquad\qquad &  r_{20} =    2 w (4+41 w)\, ,  \\
r_{03}  =  \frac{1}{2} \left(5-10 w+5 w^2-z\right)\, ,   &  \qquad\qquad  &  r_{12}  =  \frac{1}{2} \left(-5-30 w+35 w^2-3 z\right)\, , \\
r_{21}  =  1 - 2 w + 41 w^2 - z  \, , &\qquad\qquad  &  r_{30}  =   8 w (1+4 w) \, .
\end{array}
\ee
These two independent Picard--Fuchs operators $\mathcal{L}_1$ and $\mathcal{L}_2$ are differential operators of order two and three, as expected from the structure of the chiral ring associated to a Picard--Fuchs system of a two moduli Calabi--Yau threefold geometry \cite{Hosono:1993qy}. Some further analysis determines the discriminant locus of this Picard--Fuchs system and reveals that, in addition to the divisors related to large volume point, $z=0$ and $w=0$, there are interesting divisors defined by
\begin{gather}
(1-w)^4 - 2(1+6w+w^2)z + z^2  = 0  \ , \\
-(1-w)^8+4(1-w)^4(1-34w+w^2) z - 2(3+372w+1298w^2+372w^3+3w^4)z^2  \nonumber \\
+4(1-34w+w^2)z^3-z^4 = 0 \ .
\end{gather}
If we identify $w:=q_1$ and $z:=q_0 q_1^2$, these agree with the Coulomb branch singular loci we found in \cite{Jockers:2012zr}.

That the Coulomb branch singular loci correspond to singular divisors from the Picard--Fuchs system is not surprising \cite{summing}, but the interpretation here is quite satisfying.  Since a fundamental matrix of the Picard--Fuchs system has nontrivial branching around a singular divisor, $e^{-K}$ must be divergent along it.  As pointed out in \cite{Benini:2012, Doroud:2012}, the asymptotic form of the integrand of the partition function (\ref{eqn:Z-GN}) is the effective twisted superpotential $\widetilde{W}_{\textrm{eff}}(\sigma)$ on the Coulomb branch.  The defining feature of the Coulomb branch singular loci in FI parameter space is that only along these loci can $\widetilde{W}_{\textrm{eff}}(\sigma)$ be extremized with respect to $\sigma$.   Recall that if $\sigma=\sigma_p$ is such an extremum, then so is $\lambda \sigma_p$ for $\lambda\in\mathbb{C}$ \cite{summing,Jockers:2012zr}.  At an extremum, $\widetilde{W}_{\textrm{eff}}(\sigma_p) = 0$ and we must go to the next order in an expansion of the integrand of (\ref{eqn:Z-GN}) around large $\sigma$.  Since the next term in the expansion exhibits a power-law divergence, the portion of the $\sigma$-contour we added at infinity to enclose the poles will actually become divergent, causing the partition function to diverge, as it must if it is to be identified with $e^{-K}$.

\section{Conclusions and future directions}

In this paper, we argued that the two-sphere partition functions for GLSMs that flow to Calabi--Yau NLSMs, calculated using the localization technique in \cite{Benini:2012, Doroud:2012}, computes the K\"ahler potential of the quantum K\"ahler moduli space of the Calabi--Yau threefold via
\begin{equation}
Z_{S^2}(\textrm{GLSM}) = \exp \big(-K(\textrm{CY}_3)\big) \, .
\end{equation}
We verified this for the quintic, corresponding to an abelian GLSM \cite{Witten:1993yc}, and for \rodland's Pfaffian Calabi--Yau threefold in $\mathbb{P}^6$, corresponding to a non-abelian GLSM \cite{Hori:2006dk, Hori:2011pd}.  Heartened by an exact matching with results known from mirror symmetry, we then 
studied a non-complete intersection Calabi--Yau threefold for which no mirror is currently known.  We conjectured the K\"ahler potential of the quantum K\"ahler moduli space of the Gulliksen--Neg\aa{}rd determinantal Calabi--Yau threefold in $\mathbb{P}^7$, verifying that the results are consistent with certain known geometric quantities as well as a GLSM computation of Coulomb branch singular loci from \cite{Jockers:2012zr}.  These agreements led us to conjecture genus zero Gromov--Witten invariants that, to our knowledge, have not been computed in the literature.  We also noticed that the exact partition function on the two-sphere can be written in a factorized form, hinting at a relationship to topological-anti-topological fusion \cite{Cecotti:1991me}, which may also suggest an avenue by which our conjecture could be proven.

We should emphasize that the partition function is valid anywhere in the K\"ahler moduli space.  In this note we have focussed on extracting Gromov--Witten invariants by expanding the partition function in the vicinity of a large volume point. However, we believe that our techniques are applicable more generally, allowing one to extract invariants in the vicinity of other special points in the moduli space. For instance, we expect that the expansion about an orbifold point in the quantum K\"ahler moduli space computes orbifold Gromov--Witten invariants \cite{Coates:2007,Bayer:2007fk,Bouchard:2007ys}, see Appendix \ref{sec:quintic-calc} for the quintic partition function expanded around its Landau--Ginzburg orbifold point.

These same techniques can be applied to any Calabi--Yau manifold for which a GLSM is known (and for which the FI parameters provide a rational basis of the K\"ahler cone), apparently converting the problem of computing Gromov--Witten invariants into the problem of evaluating Barnes integrals (see \cite{math.AG/9912109} for earlier work along these lines) --- a problem that is significantly simpler, at least in these examples.  It is also worth noting that the expression for the fundamental period arising from non-abelian GLSMs that was conjectured by Hori and Vafa in \cite{Hori:2000kt}, bears a striking resemblance to the Barnes integrals we have studied here, with Gamma functions expressed via their integral representations.\footnote{We thank C.~Vafa for pointing this out to us.}  We hope to elucidate these connections in future work.

While our observations allow one to study the K\"ahler moduli space of a Calabi--Yau manifold (admitting a GLSM realization) without the crutch that is mirror symmetry, it would be exciting to take this one step further and use the methods to construct the mirror.  The localization method of \cite{Benini:2012, Doroud:2012} was geared towards $A$-model data, so it would be interesting to study whether there is an analog of the localization method for the $B$-model (either by coupling to a background gauge field for the axial R-symmetry, or by working with the same background fields as before but choosing twisted chirals and twisted vectors for matter fields).  Knowing the partition function for these ``$B$-localized'' models and matching to an ``$A$-localized'' partition function could allow one to infer a mirror GLSM and, possibly, mirror manifold.

Finally, connecting to string theory, for type II compactifications, our results have the striking implication that one can obtain the exact-in-$\alpha'$ spacetime K\"ahler potential and prepotential for all the geometric moduli fields: for the K\"ahler moduli using the methods we have outlined, and for the complex structure moduli using standard $B$-model computations.\footnote{Note that in spacetime, some of these fields become quaternionic and are subject to further $g_s$ corrections.}  For heterotic theories, the geometric moduli will mix with bundle moduli even around standard embedding, suggesting an intriguing direction for generalization by exploring whether these localization techniques can be applied to $(0,2)$ theories.

\bigskip
\subsection*{Acknowledgments}

We would like to thank
Christopher Beem,
Francesco Benini,
Tom Faulkner,
Jaume Gomis,
Simeon Hellerman,
Kentaro Hori,
Nabil Iqbal,
Bumsig Kim,
Albrecht Klemm,
Johanna Knapp,
Peter Koroteev,
Sungjay Lee,
Peter Mayr,
Ronen Plesser,
Joe Polchinski,
Kevin Schaefer,
Frank-Olaf Schreyer,
Eric Sharpe,
Samson Shatashvili,
and
Cumrun Vafa,
for useful discussions and correspondence.
D.R.M. thanks the Aspen Center for Physics and the Kavli Institute for the Physics and Mathematics of the Universe for hospitality, and V.K. and D.R.M thank the Simons Center for Geometry and Theoretical Physics for hospitality during the 2012 Summer Simons Workshop in Mathematics and Physics.
H.J.~is supported by the DFG grant KL 2271/1-1;
V.K.~is supported in part by the National Science Foundation under Grant No. PHY11-25915;
J.M.L.~is supported by the National Science and Engineering Research Council of Canada;
D.R.M.~is supported in part by NSF Grants DMS-1007414 and PHY-1066293.
M.R.~is supported by World Premier International Research Center
Initiative (WPI Initiative), MEXT, Japan. 

\bigskip
\goodbreak

\appendix

\section{Partition function of the quintic threefold}
\label{sec:quintic-calc}

The integral we need to evaluate is
\begin{equation}
Z_{\textrm{hyp}} = e^{4\pi  \rr r}\sum_{m\in \Z} e^{-i \theta m} \int_{\rr-i\infty}^{\rr+i\infty} \frac{d\tau}{2\pi i} \ e^{4\pi  r \tau} (Z_\Phi)^n \, Z_P \, ,
\end{equation}
where $0<\rr<\frac{1}{n}$, and
\begin{equation}
Z_\Phi := \frac{\Gamma\big(\tau - \frac{m}{2}\big)}{\Gamma\big(1-\tau-\frac{m}{2}\big)} \, , \qquad Z_P := \frac{\Gamma\big(1-n\tau + \frac{n m}{2}\big)}{\Gamma\big(n\tau +  \frac{n m}{2}\big)} \, .
\end{equation}
For $r\gg0$, we close the contour in the left-halfplane, so only the poles in $Z_\Phi$ will contribute to the residue and are located at\footnote{Here we see that if we had not restricted to $0<\rr<\frac{1}{n}$, we would either not encircle all of poles from $Z_\Phi$, or we would encircle all of the $Z_\Phi$ and some of the $Z_P$ poles.}
\begin{equation}
\tau = \tau_p(m,k) := \frac{m}{2}-k\, , \qquad k \geq \max\{0,m\} ~~\Longrightarrow ~~ m \leq k \, .
\end{equation}
Near each pole, $\tau_p$, we can write $\tau = \tau_p + \epsilon$ and write the contribution from that particular pole as a contour integral over $\epsilon$, with the contour chosen to enclose only the pole at $\epsilon=0$.  The partition function can thus be expressed as
\bea
Z_{\textrm{hyp}} &=& \sum_{k=0}^\infty \sum_{m\leq k} e^{-im\theta} \oint \frac{d\epsilon}{2\pi i} e^{4\pi r (\tau_p+\epsilon-\rr)}\, \frac{\Gamma\big(\tau_p+\epsilon-\frac{m}{2}\big)^n}{\Gamma\big(1-\tau_p-\epsilon-\frac{m}{2}\big)^n}\  \frac{\Gamma\big(1-n\tau_p-n\epsilon+\frac{nm}{2}\big)}{\Gamma\big(n\tau_p+n\epsilon+\frac{nm}{2}\big)}    \nonumber \\
&=& \sum_{k=0}^\infty \sum_{m\leq k} e^{-im\theta} \oint \frac{d\epsilon}{2\pi i} e^{4\pi r (\frac{m}{2}-k+\epsilon-\rr)}\, \frac{\Gamma(-k+\epsilon)^n}{\Gamma(1+k-\epsilon-m)^n} \ \frac{\Gamma(1+nk-n\epsilon)}{\Gamma(-nk+n\epsilon+nm)} \, .
\eea
After switching the order of integration and summation,
changing the summation variable $m \rightarrow l := k-m$, using the gamma function identity $\Gamma(x)\Gamma(1-x)\sin(\pi x) = \pi$, and defining $z:=\exp(-2\pi r + i\theta)$, the partition function becomes
\begin{gather}
Z_{\textrm{hyp}} =  \oint \frac{d\epsilon}{2\pi i}\   (z\zb)^{\rr-\epsilon} \ \frac{\pi^{n-1} \sin(n\pi\epsilon) }{\sin^n (\pi \epsilon)} \left\vert \sum_{k=0}^\infty  (-1)^{nk}\, z^k\
 \frac{\Gamma(1+nk-n\epsilon) }{\Gamma(1+k-\epsilon)^n}\right\vert^2 \, ,
\end{gather}
where complex conjugation does not act on $\epsilon$.

Note that one can similarly expand around the Landau-Ginzburg orbifold point, where $r\ll 0$, obtaining (we have set $\rr =\frac{1}{n}$)
\bea
Z_{\textrm{LG}} &=& \,   \frac{1}{n}\sum_{\delta=0}^{n-2} (-1)^\delta (z\zb)^{-\frac{\delta}{n}} \frac{  \Gamma (\frac{1+\delta}{n} )^n }{\Gamma(\delta+1)^2 \Gamma (\frac{n-1-\delta}{n} )^n }  \, \nonumber \\
 && \qquad \times \left\vert     {}_{n-1}F_{n-2} \left( \Big\{\tfrac{1+\delta}{n}, \cdots, \tfrac{1+\delta}{n} \Big\}  ; \   \Big\{ \tfrac{2+\delta}{n}, \cdots, \widehat{\tfrac{(n-\delta) + \delta}{n}} \cdots, \tfrac{n+\delta}{n} \Big\}\, ; \, \frac{(-1)^n}{n^n z} \right) \right\vert^2   \, , \qquad
\eea
where~$\widehat{\phantom{\cdots}}$~indicates that one should omit the term beneath it. This expression exactly matches the K\"ahler potential extracted from the mirror symmetry calculation in \cite{Candelas:1990rm}.
In general, we expect to be able to expand around any singular divisor associated with an asymptotic region in FI parameter space in the associated GLSM.

\section{Lines in the Gulliksen--Neg\aa{}rd Calabi--Yau threefold}
\label{sec:lines}

As discussed in Section~\ref{sec:GWinv} and as shown in~\cite{Jockers:2012zr}, there are $N_{C_1}=N_{C_0}=56$ lines in the homology classes $[C_1]$ and $[C_0]$ of the generic GN Calabi--Yau threefold $Y$, which we identify with the genus zero Gromov--Witten invariants $\tilde N_{1,0}$ and $\tilde N_{0,1}$ of $Y$.  In this Appendix, we enumerate the number of lines $N_\Gamma$ of the homology class $[\Gamma]$ corresponding to the genus zero Gromov--Witten invariant $\tilde N_{\frac12,\frac12}$ of Section~\ref{sec:GWinv}. We use standard tools of algebraic geometry and the Schubert Calculus as explained, for instance, in \cite{GriffithsHarris1994,Fulton1997,BottTu2010}.

In order to enumerate these lines, we describe the GN threefold by the incidence correspondence \eqref{GNCY}, which realizes the GN Calabi--Yau threefold $Y$ as the zero locus in $\P^7 \times G(2,4)$ of a generic global holomorphic section $f$ of the rank eight bundle $\cO(1)^{\oplus 4}_{\P^7} \otimes \cU^*_{G(2,4)}$ (where $\cU^*_{G(2,4)}$ denotes the dual of the rank two universal subbundle of $G(2,4)$), i.e.,
\begin{equation} \label{eq:GNzerolocus}
    Y \,=\, \left\{\ (\phi,p)\,\in\,\P^7 \times G(2,4) \ \middle| \ f(\phi,p)\,=\,\left( f_1(\phi,p),\ldots, f_4(\phi,p) \right)\,=\, 0 \ \right\} \ ,
\end{equation}
where we decomposed $f$ into four sections $f_1, \ldots, f_4$ of the rank two bundle $\cF = \cO(1)_{\P^7} \otimes \cU^*_{G(2,4)}$.

In the ambient space $\P^7 \times G(2,4)$ of $Y$, the lines in the homology class $[\Gamma]$ have bi-degree~$(1,1)$ with respect to $\P^7$ and $G(2,4)$. Thus, we first describe the moduli space $\cM_{1,1}$ of bi-degree~$(1,1)$ lines in the ambient space $\P^7 \times G(2,4)$, then we enumerate the number of lines $N_\Gamma$ by restricting the moduli space $\cM_{1,1}$ to the zero locus of a generic section $f$.

First, we construct an auxiliary variety $\cV_{\P^1\times\P^1}$ as the product of the moduli spaces of $\P^1$s in $\P^7$ and $G(2,4)$, respectively --- i.e., $\cV_{\P^1\times\P^1}$ consists of $\P^1\times\P^1$  embedded diagonally in $\P^7\times G(2,4)$. The moduli space of $\P^1$s in $\P^7$ is given by all 2-planes in $\C^8$, which is just the Grassmannian $G(2,8)$. As in Section~\ref{sec:GWinv}, the moduli space of $\P^1$s in $G(2,4)$ consists of the set of all hyperplanes $V \subset \C^4$ together with a line $W \subset V$. The hyperplanes $V$ are parametrized by the Grassmannian $G(3,4)$, while the line $W$ corresponds to a point in $\P^2[V]$. Therefore, the moduli space of $\P^1$s in $G(2,4)$ is expressible as the fibration $\P^2[\cU_{G(3,4)}] \rightarrow G(3,4)$ in terms of the rank three universal subbundle $\cU_{G(3,4)}$ of $G(3,4)$.  We arrive at the auxiliary variety
\begin{equation} \label{eq:VP1P1}
  \cV_{\P^1\times\P^1}\,=\, G(2,8) \times \left( \begin{CD} \P^2[\cU_{G(3,4)}] \\ @VVV \\ G(3,4) \end{CD} \right) \ , \qquad \dim_\C \cV_{\P^1\times\P^1} \,=\,17 \ .
\end{equation}  

Next, we turn to the moduli space $\cM_{1,1}$. The zero locus of a (generic) global section of the hyperplane line bundle $\cO(1,1)_{\P^1\times\P^1}$ yields a line of bi-degree $(1,1)$ in $\P^1\times\P^1$, and the moduli space of lines of bi-degree $(1,1)$ in $\P^1\times\P^1$ is the projective space of hyperplane sections $\P^3[ {\rm Hom}(\cO(-1,-1)_{\P^1\times\P^1},\cO)]$.\footnote{%
Actually, a non-zero global hyperplane section of $\cO(1,1)_{\P^1\times\P^1}$ that factorizes into sections of $\cO(1,0)_{\P^1\times\P^1}$ and $\cO(0,1)_{\P^1\times\P^1}$ gives rise to a reducible curve of two $\P^1$s of degree $(1,0)$ and $(0,1)$ touching at a common point. However, these non-generic hyperplane sections arise only in co-dimension one in the projective space $\P^3[ {\rm Hom}(\cO(-1,-1)_{\P^1\times\P^1},\cO)]$, so these degenerate lines will not appear in a generic GN threefold $Y$.}
We engineer the moduli space $\cM_{1,1}$ as a fibration over $\cV_{\P^1\times\P^1}$, where the fiber over each point $\{\P^1 \times \P^1\} \in \cV_{\P^1 \times \P^1}$ is its projective space of hyperplane sections.  Since the lines in the 2-planes of $G(2,8)$ are the points of $\P^1\subset\P^7$, and since the 2-planes $\Pi$ with $W\subset\Pi\subset V$ for $(V,W) \in \left( \P^2[\cU_{G(3,4)}] \rightarrow G(3,4)\right)$ are the points of $\P^1\subset G(2,4)$, the bundle of hyperplane sections is the tensor product $\cU^*_{G(2,8)} \otimes \cQ^*_{\P^2[U_{G(3,4)}]}$ of the dual rank two universal subbundle of $G(2,8)$ and the dual rank two universal quotient bundle of $\P^2[U_{G(3,4)}]$. Altogether, the moduli space $\cM_{1,1}$ of lines of bi-degree $(1,1)$ in the ambient space $\P^7\times G(2,4)$ becomes
\begin{equation} \label{eq:M11}
   \cM_{1,1}\,=\,\begin{CD} \P^3\left[ \cU^*_{G(2,8)} \otimes \cQ^*_{\P^2[U_{G(3,4)}]} \right] \\ @VVV \\ \cV_{\P^1\times\P^1} \end{CD} \ , \qquad \dim_\C \cM_{1,1} \,=\, 20 \ .
\end{equation}

In the following, we will need the cohomology ring of the variety $\cM_{1,1}$, which can be describes by standard techniques in algebraic geometry. The cohomology ring $H^*(G(2,8),\Z)$ is generated by the Schubert classes $\sigma_1$, $\sigma_2$, and $\sigma_3$ of degree two, four and six; the cohomology ring $H^*(G(3,4),\Z)$ is generated by the Schubert class $x$ of degree two; and the cohomology rings of the projective fibers $\P^2[\cU_{G(3,4)}]$ and $\P^3\left[ \cU^*_{G(2,8)} \otimes \cQ^*_{\P^2[U_{G(3,4)}]} \right]$ are generated by the hyperplane classes $y$ and $\xi$, each of degree two.  For our purposes, the relevant relations among the Schubert classes $\sigma_1$ and $\sigma_2$ are
\begin{equation} \label{eq:Sch}
\begin{aligned}
    &\sigma_1^{12}\,=\,132\, \sigma_{\{6,6\}} \ , &\quad &\sigma_1^{10} \sigma_2\,=\,90\,\sigma_{\{6,6\}} \ , &\quad &\sigma_1^8\sigma_2^2\,=\,62\, \sigma_{\{6,6\}}\ ,
           &\quad &\sigma_1^6\sigma_2^3\,=\,43\,\sigma_{\{6,6\}} \ ,  \\
    &\sigma_1^4\sigma_2^4\,=\, 30\,\sigma_{\{6,6\}} \ , &\quad &    \sigma_1^2\sigma_2^5\,=\,21\,\sigma_{\{6,6\}} \ , &\quad & \sigma_2^6\,=\,15\,\sigma_{\{6,6\}} \ ,
\end{aligned}
\end{equation}
where $\sigma_{\{6,6\}}$ denotes the class of a point in $G(2,8)$.  Furthermore, from the total Chern classes of the bundles $\cU_{G(3,4)}$ and $\cU^*_{G(2,8)} \otimes \cQ^*_{\P^2[U_{G(3,4)}]}$, we deduce --- using the fibrational structures of the two projective bundles in $\cV_{\P^1\times\P^1}$ and $\cM_{1,1}$ --- two additional relations
\begin{equation} \label{eq:CohRel}
\begin{aligned}
  0 &\,=\,y^3 - y^2\,x + y\,x^2 - x^3 \ , \\
  0 &\,=\,\xi^4 + \xi^3 ( 2 \sigma_1+2 x-2 y ) + \xi^2(3 \sigma_1^2-2 \sigma_2+3 x^2+3 \sigma_1 x-4 x y+3 y^2-3 \sigma_1 y) \\
  &\quad+ \xi \left(2 \sigma _1^3-2 \sigma _2 \sigma _1+3 \sigma _1 x^2-2 x^2 y+3 \sigma _1^2 x-2 \sigma _2 x
          +2 x y^2-4 \sigma _1 x y+3 \sigma _1 y^2-3 \sigma _1^2 y+2 \sigma _2 y \right)\\
  &\quad\quad + \left( \sigma _1^4-2 \sigma _2 \sigma _1^2+\sigma _2^2+\sigma _2 x^2+x^2 y^2-\sigma _1 x^2 y+
          \sigma _1^3 x-\sigma _2 \sigma _1 x+\sigma _1 x y^2-\sigma _1^2 x y \right. \\
   &\hspace{4in}\left. +\sigma _2 y^2-\sigma _1^3 y+\sigma _2 \sigma _1 y  \right) \ .
\end{aligned}    
\end{equation}
Finally, we note that $\xi^3\,y^2\,x^3\,\sigma_{\{6,6\}}$ is the class of a point in the variety $\cM_{1,1}$, i.e.,
\begin{equation} \label{eq:Pint}
   \int_{\cM_{1,1}} \xi^3\wedge y^2\wedge x^3\wedge \sigma_{\{6,6\}} \,=\, 1 \ .
\end{equation}   

With all these ingredients at hand, we are ready to enumerate the number $N_\Gamma$ of lines in the GN threefold $Y$. The holomorphic sections $f_k$ of the rank two bundle $\cF$ appearing in the incidence correspondence \eqref{eq:GNzerolocus} induce holomorphic sections $\tilde f_k$ of the rank six bundle
\begin{equation} \label{eq:wF}
  \widetilde\cF \,=\, \cU^*_{G(2,8)} \otimes \cU^*_{G(3,4)} \,\simeq \, \cU^*_{G(2,8)} \otimes  \cO(1)_{\P^2[\cU_{G(3,4)}]} \oplus \cU^*_{G(2,8)} \otimes \cQ^*_{\P^2[\cU_{G(3,4)}]}  \ .
\end{equation}
The sections $\tilde f_k$ split into two components $(\tilde f_k^{(1)},\tilde f_k^{(2)})$ according to the indicated decomposition of $\widetilde\cF$.  If the sections $\tilde f_k^{(1)}$ of $\cU^*_{G(2,8)} \otimes \cO(1)_{\P^2[\cU_{G(3,4)}]}$ simultaneously vanish and if the sections $\tilde f_k^{(2)}$ of $\cU^*_{G(2,8)} \otimes  \cQ^*_{\P^2[\cU_{G(3,4)}]}$ are all proportional to one another, then a $\P^1$ of bi-degree $(1,1)$ resides in the zero locus of the rank eight bundle $\cF^{\oplus 4}$ over $\P^7 \times G(2,4)$.  The first condition ensures that the zeros of all sections $\tilde f_k^{(1)}$ describe a common $\P^1\subset G(2,4)$, while the second condition guarantees that all sections $\tilde f_k^{(2)}$ realize the same hyperplane section --- and, hence, the same bi-degree $(1,1)$ line --- over a surface $\P^1\times\P^1\subset \P^7\times G(2,4)$.  To compute the number $N_\Gamma$, we calculate the complete intersection locus in $\cM_{1,1}$ of the sections $\hat f_k$ of the rank five bundle
\begin{equation}
  \widehat\cF \,=\, \cU^*_{G(2,8)} \otimes \cO(1)_{\P^2[\cU_{G(3,4)}]} \oplus \left( \slfrac{$\cU^*_{G(2,8)} \otimes \cQ^*_{\P^2[\cU_{G(3,4)}]}$}{$\cO(-1)_{\P^3\left[ \cU^*_{G(2,8)} \otimes \cQ^*_{\P^2[U_{G(3,4)}]} \right] }$} \right) \ ,
\end{equation}
where the additional quotient in the second summand of $\widehat\cF$ accounts for the equivalence of mutually proportional hyperplane sections as a common zero of the sections $\hat f_k^{(2)}$ induced from $\tilde f_k^{(2)}$.

Since the class of the zero-locus variety of a generic global section of a holomorphic vector bundle is its top Chern class, we can compute the number $N_\Gamma$ of bi-degree $(1,1)$ lines from
\begin{equation} \label{eq:Nint}
    N_{\Gamma}\,=\,\int_{\cM_{1,1}} c_5(\widehat\cF)^4  \ .
\end{equation}
In terms of the introduced cohomology classes of $\cM_{1,1}$, the Chern class $c_5(\widehat\cF)$ becomes 
\begin{equation}
\begin{aligned}
    c_5(\widehat\cF) \,&=\, \xi ^3 \sigma _1^2-\xi ^3 \sigma _2+2 \xi ^2 \sigma _1^3-2 \xi ^2 \sigma _2 \sigma _1+3 \xi  \sigma _1^4-5 \xi  \sigma _2 \sigma _1^2+2 \xi  \sigma _2^2+2 \sigma _1^5
      -4 \sigma _2 \sigma_1^3 \\
    &\quad +2 \sigma _2^2 \sigma _1-2 \xi ^2 x^3+2 \sigma _2 x^3-2 x^3 y^2+4 \xi  x^3 y+2 \sigma _1 x^3 y+3 \xi  \sigma _1^2 x^2 \\
    &\quad-3 \xi  \sigma _2 x^2+3 \sigma _1^3 x^2-3 \sigma _2 \sigma _1 x^2-\xi  x^2 y^2-\sigma _1 x^2 y^2+2 \xi ^2 x^2 y+3 \xi  \sigma _1 x^2 y \\
    &\quad+\sigma _1^2 x^2 y+2 \xi ^2 \sigma _1^2 x-2 \xi ^2 \sigma _2 x+3 \xi  \sigma _1^3 x-3 \xi  \sigma _2 \sigma _1+x+3 \sigma _1^4 x-5 \sigma _2 \sigma _1^2 x \\
    &\quad+2 \sigma _2^2 x-\xi  \sigma _1 x y^2+\sigma _1^2 x y^2-2 \sigma _2 x y^2+2 \xi ^2 \sigma _1 x y-\xi  \sigma _1^2 x y+4 \xi  \sigma _2 x y \\
    &\quad-\sigma _1^3 x y+2 \sigma _2 \sigma _1 x y+\xi ^3 y^2+3 \xi  \sigma _1^2 y^2-5 \xi  \sigma _2 y^2+2 \sigma _1^3 y^2-3 \sigma _2 \sigma _1 y^2\\
    &\quad+\xi ^3 \sigma _1 y+2 \xi ^2 \sigma _2 y+\xi  \sigma _2 \sigma _1 y-\sigma _1^4 y+3 \sigma _2 \sigma _1^2 y-2 \sigma _2^2 y \ .
\end{aligned}
\end{equation}
We evaluate the integral \eqref{eq:Nint} using \eqref{eq:Sch}, \eqref{eq:CohRel}, and \eqref{eq:Pint}, obtaining
\begin{equation}
   N_{\Gamma}\,=\,192 \ ,
\end{equation}   
which precisely matches the genus zero Gromov--Witten invariant $\tilde N_{\frac12,\frac12}$ predicted in Section~\ref{sec:GWinv}.

\ifx\undefined\bysame
\newcommand{\bysame}{\leavevmode\hbox to3em{\hrulefill}\,}
\fi

\end{document}